\documentclass[%
 reprint,
superscriptaddress,
nofootinbib,
longbibliography,
amsmath,amssymb,
aps,
]{revtex4-1}

\pdfoutput=1

\usepackage{graphicx}
\usepackage{dcolumn}
\usepackage{bm}

\usepackage{color,colortbl}
\usepackage{units}
\usepackage{tikz}
\usetikzlibrary{arrows,shapes}

\usepackage{algorithm}
\usepackage[noend]{algpseudocode}
\usepackage{romannum}

\usepackage{diagbox} 

\newcolumntype{C}[1]{>{\centering\let\newline\\\arraybackslash\hspace{0pt}}m{#1}}

\usepackage{subfigure}

\renewcommand{\arraystretch}{1.25}

\definecolor{mygreen}{rgb}{0.1, 0.6, 0.0}

\begin{document}
	
	\preprint{APS/123-QED}
	
	\title{Constrained multi-objective shape optimization of superconducting RF cavities considering robustness against geometric perturbations}
	
	\author{Marija Kranj\v{c}evi\'c}
	\email{marija.kranjcevic@inf.ethz.ch}
	\affiliation{Department of Computer Science, ETH~Zurich, 8092 Z\"{u}rich, Switzerland}
	
	\author{Shahnam Gorgi Zadeh}
	\altaffiliation{M.~Kranj\v{c}evi\'c and S.~Gorgi Zadeh share the first authorship. M.~Kranj\v{c}evi\'c carried out the sensitivity analysis and solved the optimization problem. S.~Gorgi Zadeh devised the problem and analyzed the results from the physics point of view.}
	\affiliation{Institute of General Electrical Engineering, University of Rostock, 18059 Rostock, Germany}
	
	\author{Andreas Adelmann}
	\affiliation{Paul Scherrer Institut (PSI), 5232 Villigen, Switzerland}%
	
	\author{Peter Arbenz}
	\affiliation{Department of Computer Science, ETH~Zurich, 8092 Z\"{u}rich, Switzerland}
	
	\author{Ursula van Rienen}
	\altaffiliation[Also at ]{Department Life, Light \& Matter, University of Rostock, 18051 Rostock, Germany}
	\affiliation{Institute of General Electrical Engineering, University of Rostock, 18059 Rostock, Germany}
	
	
	\begin{abstract}
		High current storage rings, such as the Z-pole operating mode of the FCC-ee, require accelerating cavities that are optimized with respect to both the fundamental mode and the higher order modes. Furthermore, the cavity shape needs to be robust against geometric perturbations which could, for example, arise from manufacturing inaccuracies or harsh operating conditions at cryogenic temperatures.
		This leads to a constrained multi-objective shape optimization problem which is computationally expensive even for axisymmetric cavity shapes.
		In order to decrease the computation cost, a global sensitivity analysis is performed and its results are used to reduce the search space and redefine the objective functions.
		A massively parallel implementation of an evolutionary algorithm, combined with a fast axisymmetric Maxwell eigensolver and a frequency-tuning method is used to find an approximation of the Pareto front.
		The computed Pareto front approximation and a cavity shape with desired properties are shown. Further, the approach is generalized and applied to another type of cavity.
	\end{abstract}
	
	\pacs{Valid PACS appear here}
	\maketitle
	
	
	\section{Introduction}\label{sec:introduction}
	
	Accelerating cavities are metallic chambers with a resonating electromagnetic field that are used to impart energy to charged particles in many particle accelerators. Over the last few decades, a lot of research has been carried out on improving the shape and material properties of accelerating cavities. Many attractive features of superconducting radio frequency (RF) cavities, such as high intrinsic quality factor, have made them a favorable choice in applications where high continuous wave voltage is required~\cite{Padamse1998}. The shape of the cavity, on the other hand, determines many figures of merit, such as the normalized peak electric and magnetic field on the surface of the cavity ($E_{\mathrm{pk}}/E_{\mathrm{acc}}$ and $B_{\mathrm{pk}}/E_{\mathrm{acc}}$, respectively), geometric shunt impedance ($R/Q$), cell-to-cell coupling, etc. Therefore, based on the requirements of each specific accelerator, the shape of the cavity needs to be carefully optimized.
	
	In many accelerators it is desirable to achieve a high accelerating field ($E_{\mathrm{acc}}$) in order to improve efficiency and save on equipment cost. In superconducting RF cavities the maximum achievable $E_{\mathrm{acc}}$ is limited by the maximum electric and magnetic field on the surface of the cavity. Thus, many cavity optimization methods have focused on minimizing $E_{\mathrm{pk}}/E_{\mathrm{acc}}$ and $B_{\mathrm{pk}}/E_{\mathrm{acc}}$  in order to provide room for increasing $E_{\mathrm{acc}}$~\cite{Valeri2009,Valery16,Marhauser18,Valery2005,Juntong13}. In addition to the maximum achievable $E_{\mathrm{acc}}$, the surface losses of the cavity should be minimized, which can be achieved by maximizing $G\cdot R/Q$ (where $G$ is the geometry factor). It was shown in~\cite{Valeri2009} that high $G\cdot R/Q$ typically goes along with low $B_{\mathrm{pk}}/E_{\mathrm{acc}}$, so either of the two can be considered in the optimization. Since various conflicting objective functions are usually optimized simultaneously, a trade-off between these objective functions has to be considered. In~\cite{Valeri2009,Valery16} $B_{\mathrm{pk}}/E_{\mathrm{acc}}$ is minimized, or $G\cdot R/Q$ maximized, while other figures of merit such as $E_{\mathrm{pk}}/E_{\mathrm{acc}}$, the wall slope angle, and the aperture radius of the cavity are kept fixed. In~\cite{Marhauser18} a Pareto front between $E_{\mathrm{pk}}/E_{\mathrm{acc}}$ and $B_{\mathrm{pk}}/E_{\mathrm{acc}}$ was obtained by analyzing more than a thousand geometries. A geometry was then selected from this Pareto front based on the machine requirements (high gradient applications favor a smaller $E_{\mathrm{pk}}/E_{\mathrm{acc}}$, while low loss applications favor a smaller $B_{\mathrm{pk}}/E_{\mathrm{acc}}$). 
	
	The optimization methods proposed in the literature mainly focus on optimizing the inner cells of a multi-cell cavity. The end half-cells are then optimized to get high field flatness in the cavity and ease the higher order mode (HOM) damping~\cite{Valeri2009}. The optimization of single-cell cavities is slightly different than that of multi-cell cavities. The length of the inner cells of a multi-cell cavity is usually fixed to $\beta\lambda/2$, where $\lambda$ is the wavelength of the fundamental mode (FM), and $\beta$ is the ratio of the particle velocity to the speed of light ($\beta\approx1$ for particle velocities close to the speed of light). This restriction improves acceleration, since, as a particle traverses a cell, the direction of the electric field changes and the particle receives a force in the same direction. 
	In single-cell cavities, the particle passes through only one cell, so there is no such restriction on the length of the cavity. Consequently, in single-cell optimization there is one more degree of freedom. Furthermore, since part of the field leaks into the beam pipe, in single-cell optimization the cell and the beam pipe have to be simulated together. This is in contrast to the optimization of the inner cells of multi-cell cavities, where only one cell, with appropriate boundary conditions, is considered.
	
	Many future accelerators, such as the Future Circular Collider (FCC), aim at colliding beams with unprecedented luminosities~\cite{Benedikt16}. The FCC design study includes three colliders: a hadron collider (FCC-hh), a lepton collider (FCC-ee) and a lepton-hadron collider. The aim of FCC-ee is to study properties of Z, W and Higgs bosons, as well as top quark, 
	with collision energies ranging from \unit[90]{GeV} to \unit[365]{GeV}. The high luminosity requirements demand a significant increase in the beam current of the machine, which in turn increases the HOM power deposited into the cavities by the traversing beam. Preliminary studies have suggested using single-cell cavities for FCC-hh and the Z-pole operating mode of FCC-ee (FCC-ee-Z)~\cite{FCCCDR,Zadeh18}. In the case of FCC-ee-Z, the main reasons for selecting a single-cell cavity are the issues related to the beam instability and high HOM power. In the FCC-ee-Z, the operating $E_{\mathrm{acc}}$ is a few \unit{MV/m}~\cite{FCCCDR,Zadeh18,Brunner17}. Reaching higher $E_{\mathrm{acc}}$ is precluded due to limitations on the fundamental power coupler in providing high input power per cavity (e.g., for FCC-ee-Z this is around \unit[1]{MW} per cavity). In such low $E_{\mathrm{acc}}$ and high-current operations, minimizing surface peak fields should not be the primary goal. Instead, other figures of merit, in particular the HOM aspects, have to be taken into account right from the early design stages of the cavity. Enlarging the beam pipe radius is a common approach to untrap many HOMs and reduce the loss factor. However, even with an enlarged beam pipe radius, the first dipole band usually stays trapped and cannot propagate out of the cavity~\cite{Haebel}. Additionally, the modes in the first dipole band usually have a large transverse impedance, which can cause transverse beam instability. Therefore, and in order to simplify their damping via HOM couplers, special attention should be given to the modes in the first dipole band.
	
	In addition to the RF properties of the cavity, the cavity shape needs to be robust against geometric perturbations~\cite{corno2015isogeometric,xiao2007modeling,brackebusch2015investigation,gorgizadeh2018eigenmode}, which could, for example, arise from harsh operating conditions at cryogenic temperatures or manufacturing inaccuracies. Most importantly, the frequency of the FM has to remain at its nominal value in order to avoid excessive power being fed into the cavity for maintaining $E_{\mathrm{acc}}$, and the frequencies of the modes in the first dipole band should not be very sensitive to geometric changes in order to avoid unwantedly hitting a beam spectral line or harming their damping when coaxial HOM couplers are used. 

        The major contribution of this paper lies in finding a robust cavity shape while several properties of the FM and the first dipole band are optimized at the same time. For this purpose an optimization method for constrained multi-objective shape optimization of superconducting RF cavities is proposed. The focus of optimization is on axisymmetric single-cell cavities used in high-current accelerators, e.g., FCC-ee-Z, which unlike the multi-cell cavities have not been extensively studied before. As a part of the proposed optimization approach, a global sensitivity analysis is carried out in order to quantify the relative influence of each of the geometric parameters on the figures of merit of the cavity. The novel approach makes use of the most influential parameters, reduces the search space by discarding non-influential parameters, and provides valuable insights into the optimization of RF cavities. 
	
	The structure of the paper is as follows. The parameterization of the cavity cross section and the quantities of interest are described in section~\ref{sec:QoI}. The results of the sensitivity analysis are shown in section~\ref{sec:sensitivity-analysis}, and used for search space reduction in section~\ref{sec:frequency-fixing}. The constrained multi-objective optimization problem is defined in section~\ref{sec:cmoop}, the algorithm described in section~\ref{sec:cmooa}, and the results presented in section~\ref{sec:results}. The approach described in sections~\ref{sec:QoI}--\ref{sec:cmooa} is generalized to a different type of cavity in section~\ref{sec:generalization}. Conclusions are drawn in section~\ref{sec:conclusions}.
	
	\section{Quantities of interest}\label{sec:QoI}
	
\begin{figure}[!tbh]
		\centering
		\includegraphics[width=0.8\columnwidth]{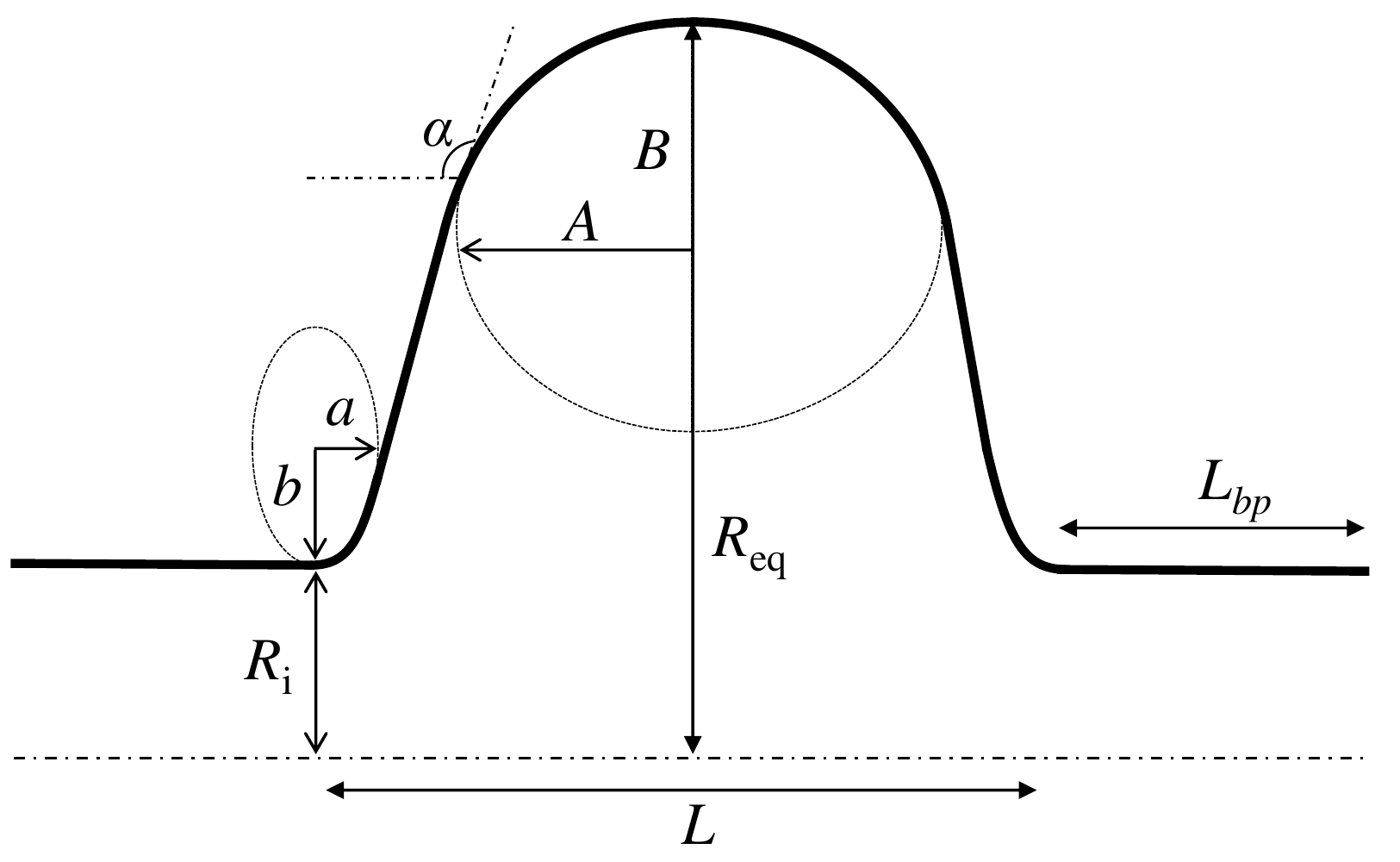}\hfill
		\caption{The parameterization of the cross section of a single-cell elliptical cavity. The variables $R_{eq}$, $L$, $A$, $B$, $a$, $b$,  and $R_i$ parameterize the cell, and $L_{bp}$ is the beam pipe length.} 
		\label{fig:SingleCell}
	\end{figure}

	Elliptical cavities are a commonly used cavity shape for the acceleration of particles with $\beta\approx1$. As shown in Fig.~\ref{fig:SingleCell}, the cross section of the cell of such cavities can be parameterized by seven variables, i.e., geometric parameters: $R_{eq}$, $L$, $A$, $B$, $a$, $b$,  and $R_i$. The wall slope angle $\alpha$ is then determined by these geometric parameters, and the variable $L_{pb}$ determines the length of the beam pipe.
	
	In constrained multi-objective shape optimization the aim is to find the geometric parameters that best satisfy the given objectives and constraints. The objectives and constraints that need to be satisfied are determined by the requirements of the accelerator. 
	The first constraint is to tune the frequency of the FM ($f$), which is typically the TM$_{010}$ mode, to a desired value. Focusing on FCC-ee-Z, in this paper $f$ is tuned to \unit[400.79]{MHz}, which is the proposed value for both FCC-ee-Z and FCC-hh. In order to have a sufficient distance for decaying the FM leaked into the beam pipe, the beam pipe length $L_{bp}$ is set to the value of the wave length, i.e., to $\lambda = \unit[748]{mm}$.
	The second constraint is $\alpha \geq 90^\circ$ in order to avoid re-entrant cavity shapes (due to the problems associated with chemical treatment).
	
	In comparison with the middle-cells of multi-cell cavities, the electric field concentration around the iris region of single-cell cavities is lower because part of the electric field leaks into the beam pipe. Due to this reason and relatively low $E_{\mathrm{acc}}$ of FCC-ee-Z, the surface peak fields are not the primary concern in the optimization.
	Lowering the surface losses helps to decrease the power released into the helium bath, which consequently reduces the amount of power required to maintain the cryogenic temperature. The surface losses of the FM are given by
		\begin{equation}
		P_c= \frac{\Big(V_{||(r=0)}\Big)^2}{G\cdot R/Q}R_s,
		\label{eq:PowerLoss}
		\end{equation}
	where $R_s$ is the surface resistance of the cavity, which depends on its material properties, $V_{||(r=0)}$ the longitudinal voltage calculated along the longitudinal axis, and $G$ and $R/Q$ the geometry factor and the geometric shunt impedance of the FM, respectively. Both $G$ and $R/Q$ depend on the shape of the cavity~\cite{Valery2005}. Eq.~(\ref{eq:PowerLoss}) indicates that $P_c$ can be minimized by maximizing $G\cdot R/Q$ (the corresponding objective function is labeled as $F_4$ in Eq.~(\ref{eq:OptimizationGoal})). 
	
\begin{figure}[!tbh]
		\centering
		\includegraphics[width=\columnwidth]{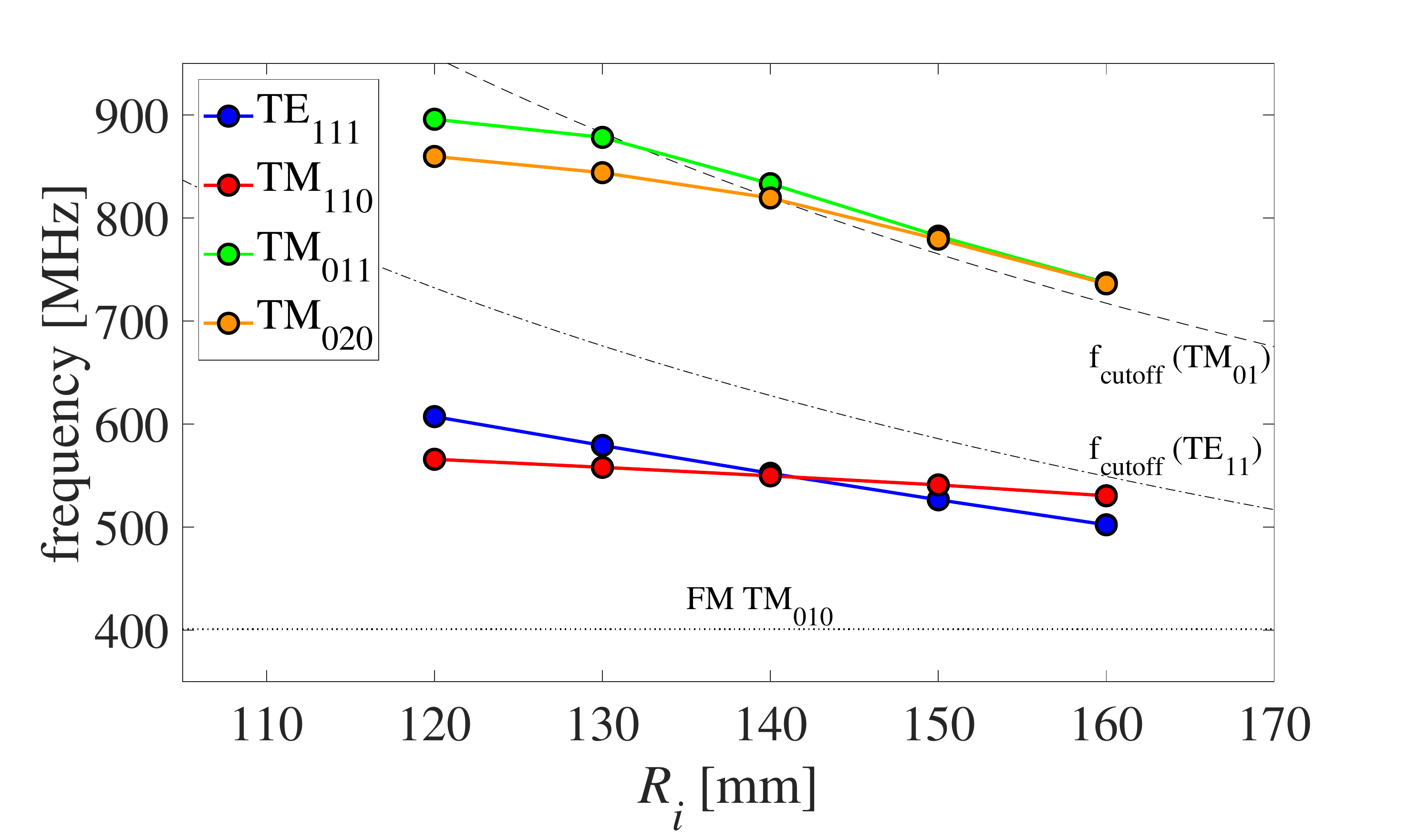}\hfill
		\caption{The dependency of the frequency of the TE$_{111}$, TM$_{110}$, TM$_{011}$ and TM$_{020}$ modes on $R_i$. For each case, the frequency of the FM, i.e., the TM$_{010}$ mode is tuned to \unit[400.79]{MHz} by varying $R_{eq}$. The parameter sweep is carried out around the geometric parameters taken from~\cite{Marija18}. An aperture radius $R_i$ approximately above \unit[145]{mm} helps to untrap the dangerous monopole HOMs. Note that the cavity in~\cite{Marija18} is optimized such that the frequencies of the TE$_{111}$ and TM$_{110}$ modes are almost equal.} 
		\label{fig:RiSweep}
	\end{figure}

	In addition to the properties of the FM, the properties of the HOM spectrum also have to be taken into account. The longitudinal loss factor is inversely proportional to the aperture radius $R_i$ of the cavity~\cite[p.~372]{Palumbo1994}. Therefore, increasing the aperture radius lowers the longitudinal loss factor and consequently the HOM power deposited into the cavity by the beam. Larger aperture radius also helps to untrap the dangerous higher order monopole modes. As shown in Fig.~\ref{fig:RiSweep}, an aperture radius roughly above \unit[145]{mm} untraps the monopole modes which typically have a large longitudinal geometric shunt impedance, i.e., the TM$_{011}$ and TM$_{020}$ modes. Therefore, a larger aperture radius is preferable (in Eq.~(\ref{eq:OptimizationGoal}), this is denoted by $F_5$). The first dipole band, however, stays trapped in the cavity, in particular the TE$_{111}$ mode whose frequency approaches the FM with enlarged $R_i$. 
	
	The damping of the HOMs is usually done using coaxial couplers, waveguide (WG) couplers, beam pipe absorbers, or their combination. Coaxial and WG HOM couplers provide better damping of trapped modes as they can be placed close to the cell. In order to allow the propagation of the HOMs into the WG, the cutoff frequency of the first mode of the WG (the TE$_{01}$ mode) has to be between the FM and the first dipole mode. If the frequency of the first dipole mode is very close to $f$, a WG coupler with larger dimensions is required, which occupies more space in the cryomodule, in particular at \unit[400.79]{MHz}. In that case the WG has to be wider in order to decrease the cutoff frequency of its TE$_{01}$ mode, and also longer in order to have a sufficient distance for decaying the FM leaked into the WG. Therefore, as another objective function, the distance between $f$ and the frequency of the first dipole mode ($f_1$), 
	which is typically the TE$_{111}$ mode, 
	has to be maximized. Since $f_1$ is larger than $f$, this is equivalent to minimizing the negative value $f - f_1$ (in Eq.~(\ref{eq:OptimizationGoal}), the corresponding objective function is labeled as $F_1$).
	
	Coaxial HOM couplers act like 3D resonant circuits that are optimized to have a notch at the FM and resonances (with a high transmission) at certain frequencies that require strong damping, such as the frequency of the TE$_{111}$ and TM$_{110}$ modes. A smaller difference between the frequencies of the first two dipole modes simplifies using a narrow band coaxial HOM coupler for damping them (such as the Hook-type coupler used in LHC cavities~\cite{Haebel1997,Sotirios2014,Roggen2015}). Therefore, another objective is to minimize the difference between the frequencies of the two trapped dipole modes (this is denoted by $F_2$ in Eq.~(\ref{eq:OptimizationGoal})). Additionally, the sum of the transverse impedances of the first two dipole modes is minimized ($F_3$ in Eq.~(\ref{eq:OptimizationGoal})). The following definition of the transverse impedance is used
		\begin{equation*}
		\frac{R}{Q}_{\perp}= \frac{\bigg(V_{||(r=r_0)}-V_{||(r=0)}\bigg)^2}{k^2r_0^2\omega U},
		\end{equation*}	
	where $k$ is the wave number, $r_0$ the offset from the axis, $\omega$ the angular frequency, and U the stored energy.
	
\begin{figure}[!tbh]
		\centering
		\includegraphics[width=\columnwidth]{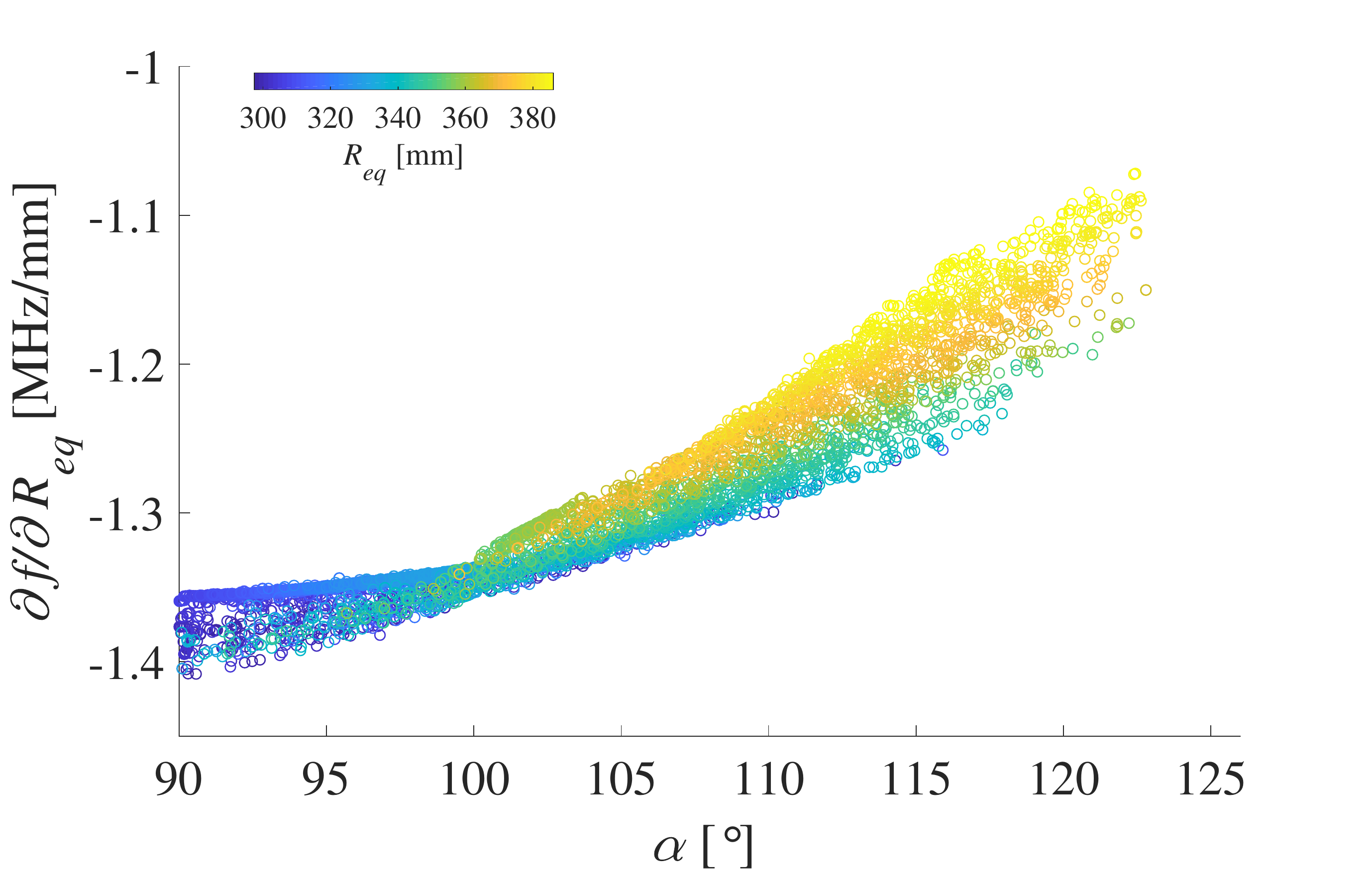}\hfill
		\caption{The local sensitivity of $f$ with respect to $R_{eq}$ plotted versus the wall slope angle $\alpha$. Around 5'000 samples are studied. The frequency of each sample is tuned to~\unit[400.79]{MHz}. The local sensitivity is calculated using the forward difference method with the step $h = \unit[1]{mm}.$ Generally, a higher wall slope angle yields a more robust $f$ against changes in $R_{eq}$.} 
		\label{fig:dfAngle}
	\end{figure}

	The optimized cavity should be robust against geometric changes that might arise due to manufacturing inaccuracies or perturbations during operation. If the FM is detuned from its nominal value, additional power has to be fed into the cavity to maintain the same $E_{\mathrm{acc}}$. Thus, tuners are used to adjust the frequency to the desired value. Frequency tuning is usually carried out by changing the length of the cells via applying a longitudinal force to the cavity~\cite[p.~431]{Padamse1998}. A large tuning force should be avoided as it might plastically deform the cavity. Therefore, the cavities have to be fabricated with a high precision to avoid large tunings afterwards. On the other hand, a cavity fabricated with a higher precision is more expensive to fabricate and handle. As shown in Fig.~\ref{fig:dfAngle}, different cavity shapes can have different sensitivities with respect to geometric perturbations. E.g., cavities with a larger wall slope angle have a more robust $f$ against changes in $R_{eq}$ (up to around \unit[30]{\%} difference between different shapes). 
	
\begin{figure}[!tbh]
		\centering
		\includegraphics[width=\columnwidth]{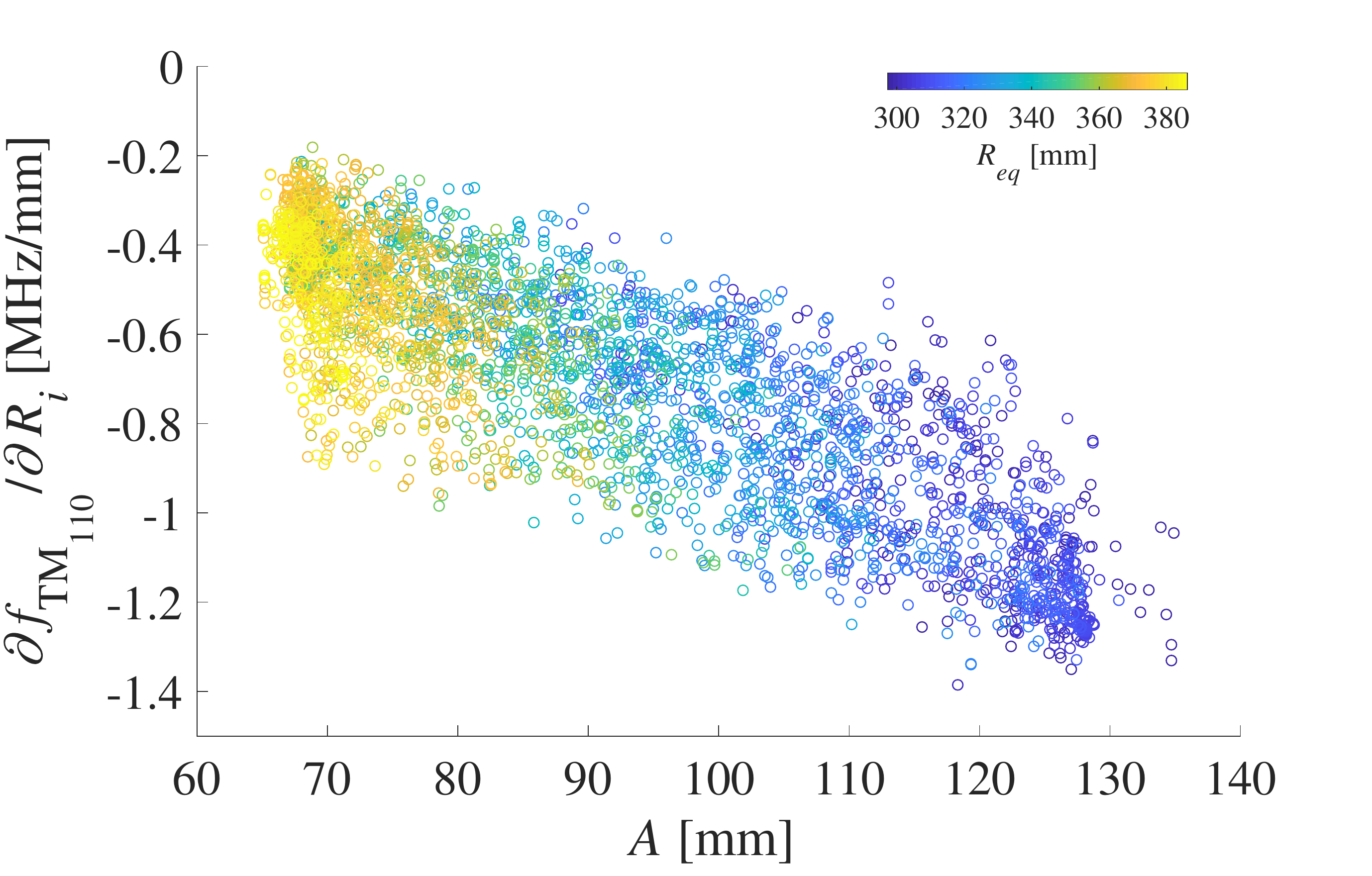}\hfill
		\caption{The local sensitivity of the frequency of the TM$_{110}$ mode with respect to $R_{i}$ plotted versus $A$. Around 5'000 samples are studied, and the frequency of the FM of each sample is tuned to~\unit[400.79]{MHz}. } 
		\label{fig:dfDipole}
	\end{figure}

	The frequency of the trapped dipole modes could also vary due to perturbations or during the tuning of the frequency of the FM. 
	Fig.~\ref{fig:dfDipole} shows the local sensitivity of the frequency of the TM$_{110}$ mode with respect to $R_i$ against the geometric variable $A$. Sensitivity of the TM$_{110}$ mode against changes in $R_i$ can vary up to a factor of seven between different shapes.
	A change in the frequency of the dipole modes could harm their damping when coaxial HOM couplers are used. Coaxial HOM couplers are usually tuned to have a resonance at the frequency of the targeted HOMs and a mismatch between the frequency of the HOM and the coaxial HOM coupler could result in a poor damping of that mode. Therefore, in addition to the robustness of $f$, the robustness of the frequencies of the first two dipole modes is of importance. 
	In Eq.~(\ref{eq:OptimizationGoal}), the objective functions corresponding to the minimization of the local sensitivities of $f$, $f_{\mathrm{TE_{111}}}$ and $f_{\mathrm{TM_{110}}}$ against geometric changes are denoted by $\hat{F}_6$, $\hat{F}_7$ and $\hat{F}_8$, respectively.
	
	To summarize, denoting by $f_2$ the frequency of the second dipole mode (which is typically the TM$_{110}$ mode) and 
		\begin{equation}
		\boldsymbol{d} = (d_1,\dots,d_7) = (R_{eq}, R_i, L, A, B, a, b),
		\label{eq:d}
		\end{equation}
	the constrained multi-objective optimization problem considered in this paper can be written as

                \begin{equation}
		\begin{aligned}
		& \underset{\boldsymbol{d}}{\text{min}}
		&& \Bigg(\underbrace{f-f_1}_{F_1},\underbrace{|f_1-f_2|}_{F_2},\underbrace{\frac{R}{Q}_{\perp1}+\frac{R}{Q}_{\perp 2}}_{F_3}, \\
		& && \underbrace{-G \cdot \frac{R}{Q}}_{F_4}, \underbrace{-R_i}_{F_5}, \underbrace{ \sum_{j}\bigg|\frac{\partial f}{\partial d_j}\bigg|}_{\hat{F}_6},\\ 
		& && \underbrace{ \sum_{j}\bigg|\frac{\partial f_{\mathrm{TE_{111}}}}{\partial d_j}\bigg|}_{\hat{F}_7}, \underbrace{ \sum_{j}\bigg|\frac{\partial f_{\mathrm{TM_{110}}}}{\partial d_j}\bigg|}_{\hat{F}_8}\Bigg), \\
		& \text{subject to} && f = 400.79 \ \text{MHz}\ \text{ and } \ \alpha[^\circ] \geq 90.
		\end{aligned}
		\label{eq:OptimizationGoal}
		\end{equation}

	In this formulation the local sensitivities of the FM and the dipole modes are calculated with respect to all seven geometric parameters, which entails a very expensive computation. In the following section a global sensitivity analysis is carried out in order to find the most influential geometric parameters on each of the frequencies. This information is then used to redefine objective functions $\hat{F}_6$, $\hat{F}_7$ and $\hat{F}_8$, reduce the search space and, consequently, reduce the computational cost of the problem.
	
	\section{Sensitivity analysis}\label{sec:sensitivity-analysis}
	In order to determine which geometric parameters have the greatest influence on $f$, $f_{\text{TE}_{111}}$ and $f_{\text{TM}_{110}}$, a variance-based global sensitivity analysis is performed~\cite{Smith:2013,uq-pam}. The geometric parameters are considered to be independent, uniformly distributed random variables. The first-order Sobol' indices and total Sobol' indices~\cite{Sobol:2001}, representing the individual and total influences of these random variables on the variance of the quantities of interest (QoIs), are computed using polynomial chaos (PC) expansion~\cite{nof-regression-pts,uqtk-1,uq-pam,schmidt2014comparison,Heller2014415}. 
	The coefficients of the PC expansion are obtained non-intrusively, because non-intrusive methods allow the use of an existing solver as a black box, requiring only an evaluation of the QoIs in a set of either deterministic or random points. Since some of the deterministic points may correspond to infeasible cavity shapes, a random sample, i.e., a collection of random points, is used. According to~\cite{nof-regression-pts}, it is enough to use a sample of size
		\begin{equation}
		(N - 1)\frac{(N+p)!}{N! \cdot p!},
		\label{eq:nof-regression-pts}
		\end{equation}
	where $N$ is the number of geometric parameters and $p$ the polynomial degree used. The values of Eq.~(\ref{eq:nof-regression-pts}), for a few relevant values of $N$ and $p$, are given in Table~\ref{Tab:nof-regression-pts}.
	
	\begin{table}[h]
		\centering
		\caption{Eq.~(\ref{eq:nof-regression-pts}) for a few relevant values of $N$ and $p$.}
		\label{Tab:nof-regression-pts}
		\begin{tabular}{|c|c|c|c|}
			\hline
			\backslashbox{$N$}{$p$} & 2 & 3 & 4 \\ \hline
			6   & 140  & 420 &  1050 \\ \hline
			7   & 216  & 720 & 1980 \\ \hline
		\end{tabular}
	\end{table}
	
	\begin{table}[h]
		\centering
		\caption{Wide intervals for the geometric parameters.}
		\label{Tab:GeometryConstraints}
		\begin{tabular}{|l|c|c|c|c|c|c|c|}
			\hline
			Variable [mm] & $R_{eq}$ & $R_i$ & $L$ & $A$ & $B$ & $a$  & $b$ \\ \hline
			Lower bound   & 325 & 145 & 240 & 65  & 65 & 10 & 10 \\ \hline
			Upper bound   & 375 & 160 & 380 & 140 & 140 & 60 & 60 \\ \hline
		\end{tabular}
	\end{table}
	
	\begin{table}[h]
		\centering
		\caption{Narrow intervals for the geometric parameters. }
		\label{Tab:int2}
		\begin{tabular}{|l|c|c|c|c|c|c|c|}
			\hline
			Variable [mm] & $R_{eq}$ & $R_i$ & $L$ & $A$ & $B$ & $a$  & $b$ \\ \hline
			Lower bound   & 345  & 145 & 280 & 70  & 70  & 45 & 45 \\ \hline
			Upper bound   & 355  & 155 & 300 & 80 & 80 & 55 & 55 \\ \hline
		\end{tabular}
	\end{table}
	
\begin{figure}
		\begin{center}
			\begin{minipage}[b]{\columnwidth}
				\centering
				\includegraphics[width=\columnwidth,clip,trim={1.45cm 5.0cm 2.3cm 0cm}]{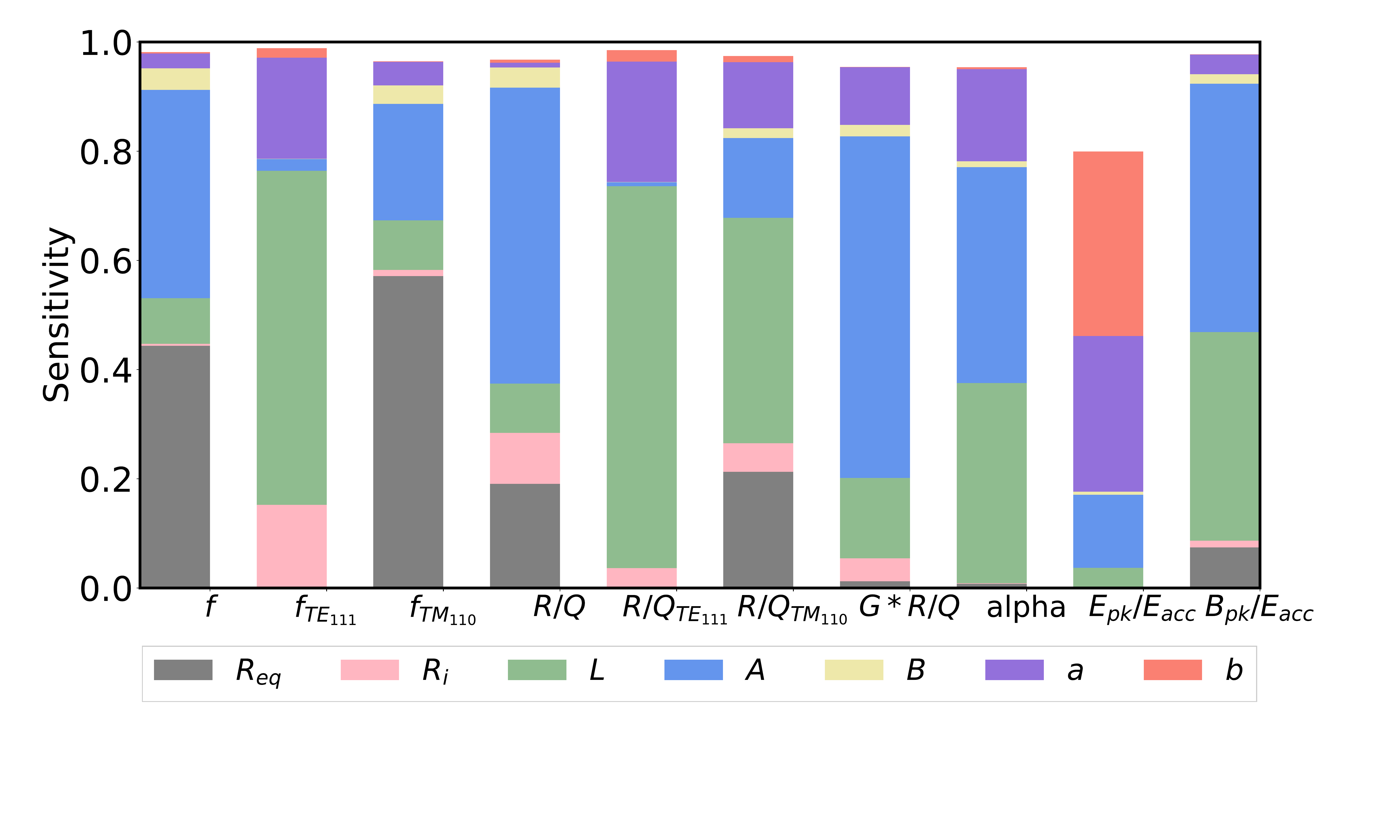}\hfill
				\caption{Main sensitivities. The intervals for the geometric parameters $R_{eq}, R_i, L, A, B, a, b$ are given in Table~\ref{Tab:GeometryConstraints}. The polynomial degree is $p=2$, so the size of the sample is 216 [cf.~Table~\ref{Tab:nof-regression-pts}].} 
				\label{fig:wide-p2-main}
			\end{minipage}
			
			\begin{minipage}[b]{\columnwidth}
				\centering
				\includegraphics[width=\columnwidth,clip,trim={1.45cm 5.0cm 2.3cm 0cm}]{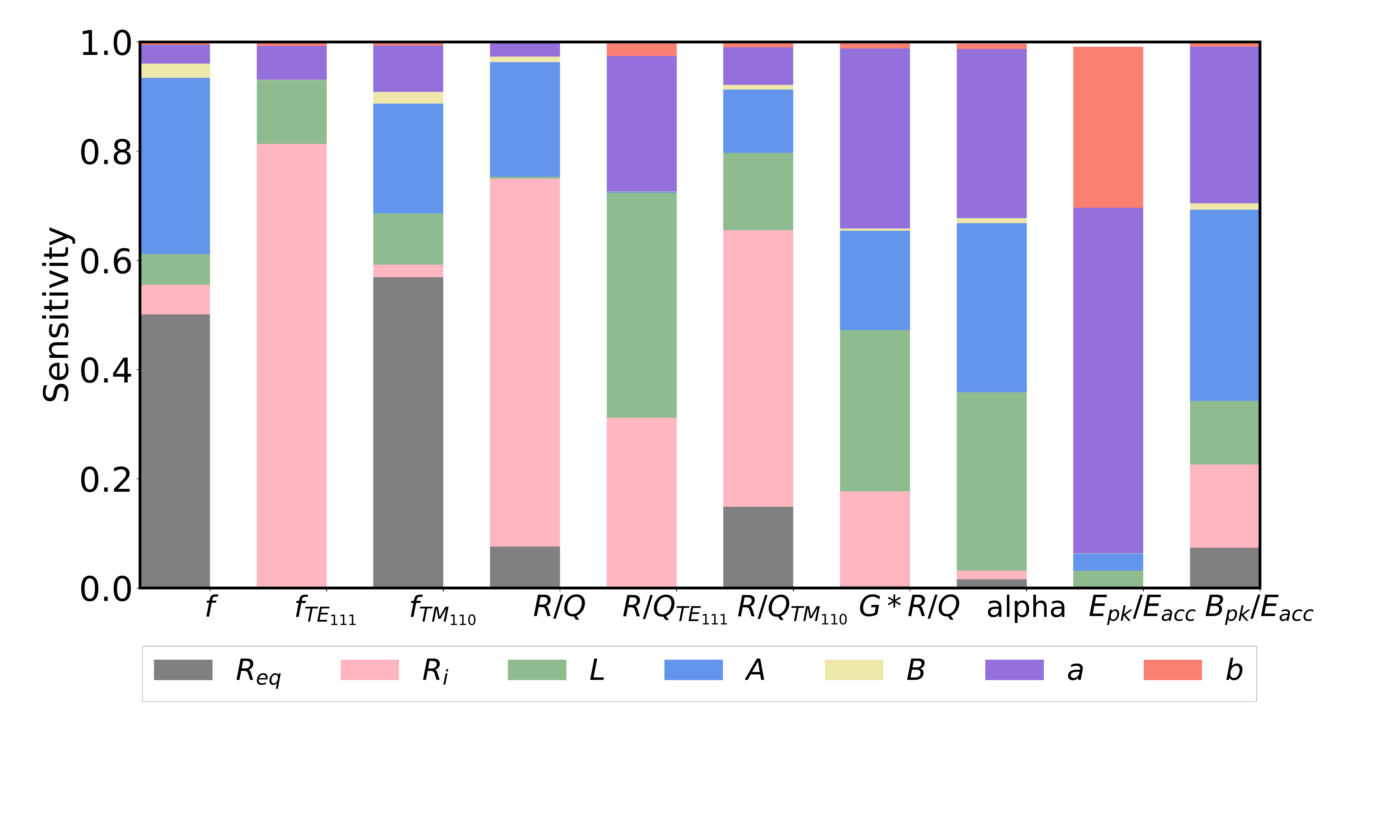}\hfill
				\caption{Main sensitivities. The intervals for the geometric parameters $R_{eq}, R_i, L, A, B, a, b$ are given in Table~\ref{Tab:int2}. The polynomial degree is $p=2$, so the size of the sample is 216 [cf.~Table~\ref{Tab:nof-regression-pts}].} 
				\label{fig:narrow-p2-main}
			\end{minipage}
		\end{center}
	\end{figure}

	For this analysis, the Uncertainty Quantification Toolkit (UQTk)~\cite{uqtk-2,uqtk-1} is used. 
	The random sample is evaluated in parallel, taking into account only feasible cavity shapes with $\alpha \geq 90^\circ$. 
	The changes in the sensitivity plots obtained for $p = 3$ (which needs 720 random points) are almost imperceptible, so only the case $p=2$ (216 random points) is shown. 
	The main sensitivities (i.e., first-order Sobol' indices), with $p=2$ and with geometric parameters in the intervals shown in Tables~\ref{Tab:GeometryConstraints} and \ref{Tab:int2}, are shown in Figs.~\ref{fig:wide-p2-main} and \ref{fig:narrow-p2-main}, respectively.
	
	These plots show the influence of the geometric parameters [cf.~Eq.~(\ref{eq:d})] on the frequencies
		\begin{equation*}
		f, f_{\text{TE}_{111}}, f_{\text{TM}_{110}},
		\end{equation*}
	but also on other QoIs
		\begin{equation*}
		\frac{R}{Q}, \frac{R}{Q}_{\perp,\mathrm{TE}_{111}}, \frac{R}{Q}_{\perp,\mathrm{TM}_{110}}, G\cdot \frac{R}{Q}, \alpha, \frac{E_{\mathrm{pk}}}{E_{\mathrm{acc}}}, \frac{B_{\mathrm{pk}}}{E_{\mathrm{acc}}}.
		\end{equation*}
	Sobol' indices are, by definition, normalized with respect to the total variance, so they (the first-order and higher-order indices) sum up to 1. Consequently, the sum of the first-order indices, representing the individual influences of the parameters, is at most 1. The higher-order indices represent mixed influences of the parameters, so the fact that the sum of the first-order indices, i.e., the height of the bars 
	in the plots, 
	is close to 1 indicates a low correlation between parameters.
	
	The geometric parameter $R_{eq}$ has the greatest influence on the frequency of the FM, $f$, (closely followed by $A$), so it is used to tune $f$ to \unit[400.79]{MHz}, i.e., to enforce the first constraint [cf.~Eq.~(\ref{eq:OptimizationGoal})]. This is explained in detail in the next section.
	
	\section{Search space reduction}\label{sec:frequency-fixing}
\begin{figure}[hb!]
		\begin{center}
			\begin{minipage}[b]{\columnwidth}
				\centering
				\includegraphics[width=\columnwidth,clip,trim={1.45cm 5.0cm 2.3cm 0cm}]{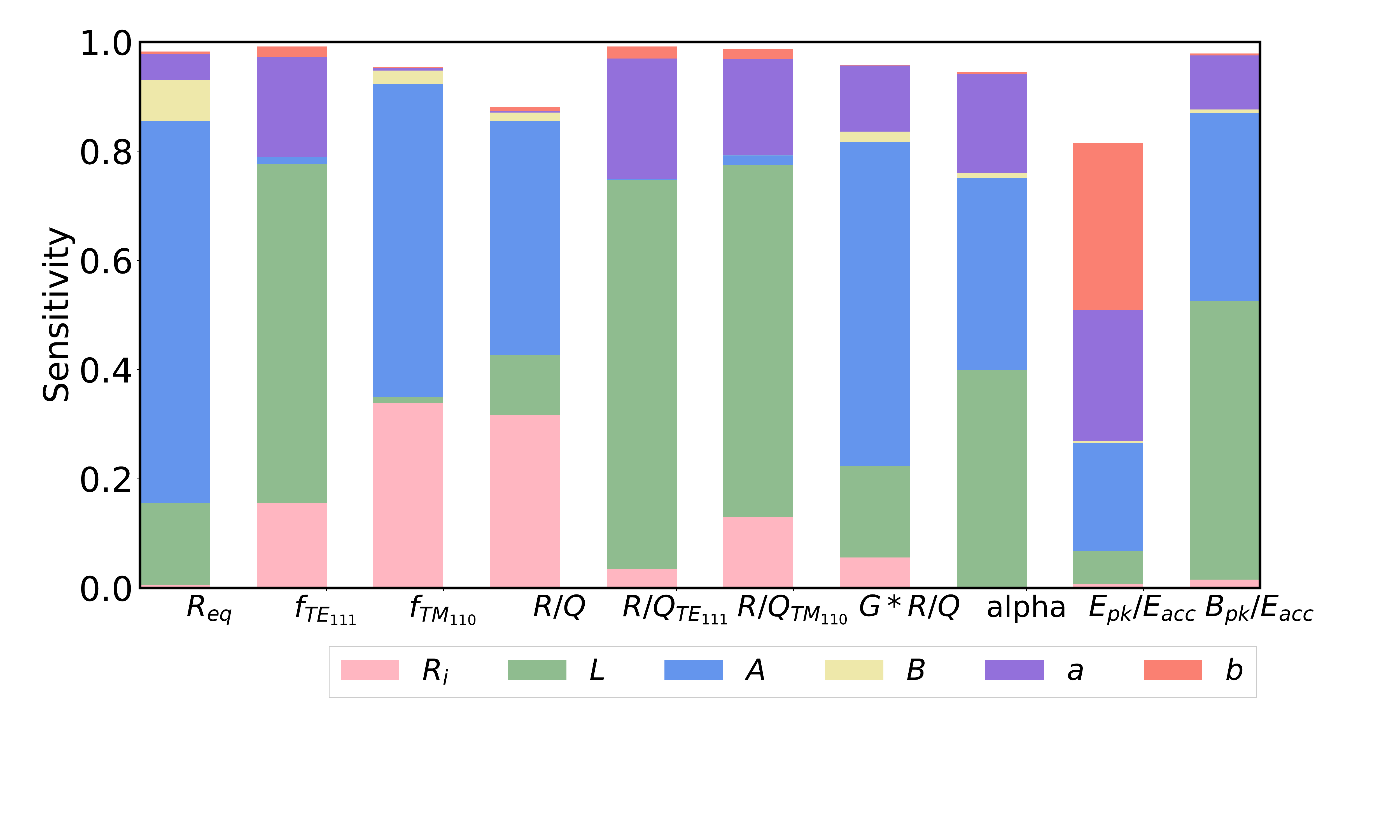}\hfill
				\caption{Main sensitivities. For a point $(R_i, L, A, B, a, b)$ inside the intervals from Table~\ref{Tab:GeometryConstraints}, $f$ is tuned to \unit[400.79]{MHz} using $R_{eq}[\text{mm}]\in[325,375]$. The polynomial degree is $p=2$, so the size of the sample is 140 [cf.~Table~\ref{Tab:nof-regression-pts}].} 
				\label{fig:fix-freq-wide-p2-main}
			\end{minipage}
			
			\begin{minipage}[b]{\columnwidth}
				\centering
				\includegraphics[width=\columnwidth,clip,trim={1.45cm 5.0cm 2.3cm 0cm}]{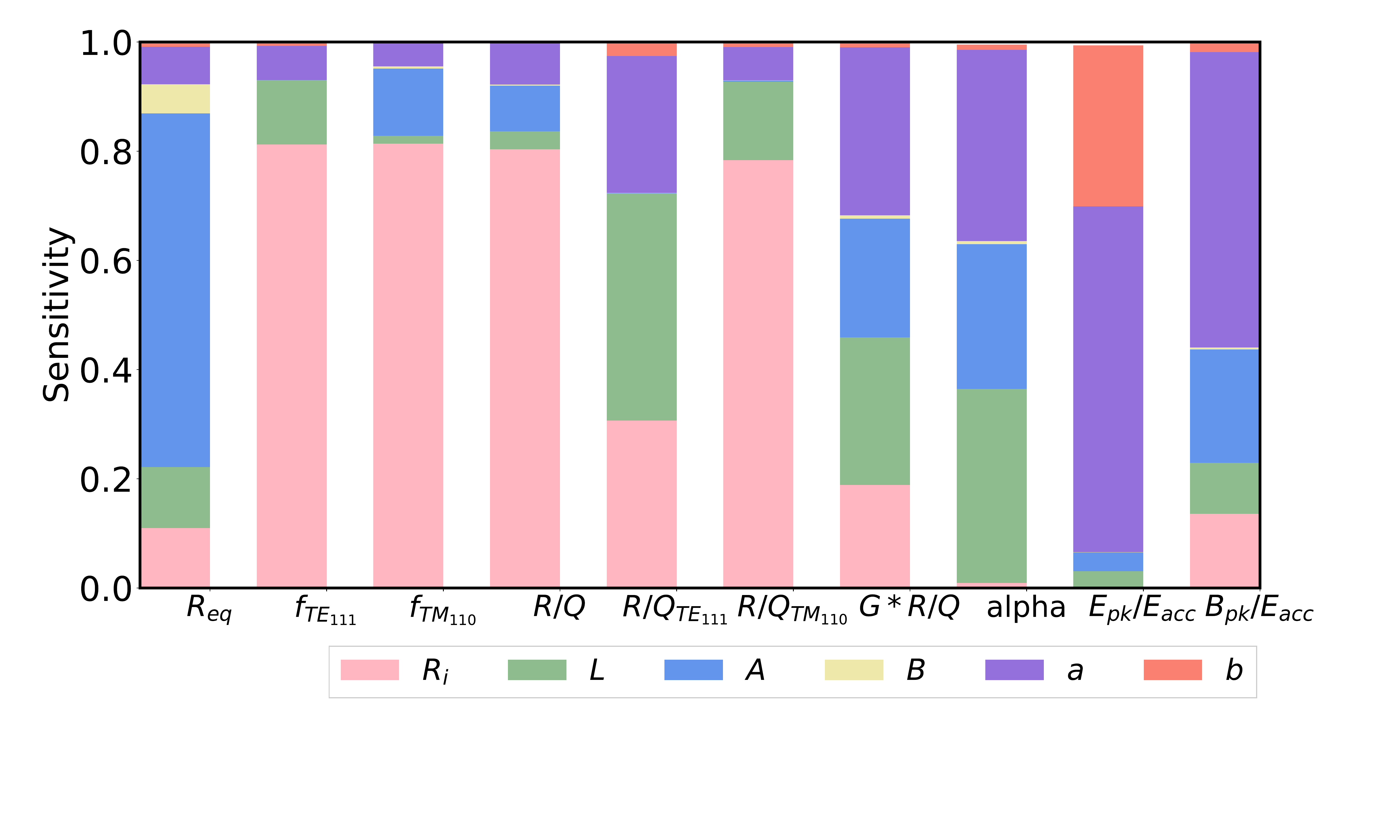}\hfill
				\caption{Main sensitivities. For a point $(R_i, L, A, B, a, b)$ inside the intervals from Table~\ref{Tab:int2}, $f$ is tuned to \unit[400.79]{MHz} using $R_{eq}[\text{mm}]\in[345,355]$. The polynomial degree is $p=2$, so the size of the sample is 140 [cf.~Table~\ref{Tab:nof-regression-pts}].} 
				\label{fig:fix-freq-narrow-p2-main}
			\end{minipage}
		\end{center}
	\end{figure}

	According to Figs.~\ref{fig:wide-p2-main} and \ref{fig:narrow-p2-main}, $R_{eq}$ has the greatest influence on $f$. Therefore, it is possible to use $R_{eq}$ to tune the frequency to \unit[400.79]{MHz}.
	Specifically, for a point 
	\[(d_2,\dots,d_7) = (R_i, L, A, B, a, b)\]
	[cf.~Eq.~(\ref{eq:d})], $d_1 =  R_{eq}[\text{mm}]\in[d_{lower},d_{upper}]$ is found such that \[f (\boldsymbol{d}) = \unit[400.79]{MHz}.\] 
	The QoIs are then computed for the cavity corresponding to $\boldsymbol{d}$ (the details are given in section~\ref{sec:optimization-algorithm} and Algorithm~\ref{alg:evaluate}).
	The main sensitivities of the QoIs and $R_{eq}$ with respect to $d_2,\dots,d_7$, considering the wide (Table~\ref{Tab:GeometryConstraints}) and narrow (Table~\ref{Tab:int2}) intervals are shown in Figs.~\ref{fig:fix-freq-wide-p2-main} and \ref{fig:fix-freq-narrow-p2-main}, respectively. The polynomial degree is again $p=2$, but the number of variables is now 6, so only 140 training points are needed.
	
	It can be seen in Figs.~\ref{fig:wide-p2-main}-\ref{fig:fix-freq-narrow-p2-main} that the influence of $B$ on all of the QoIs is very low, and that $b$ significantly influences only $E_{\mathrm{pk}}/E_{\mathrm{acc}}$. Therefore, since $E_{\mathrm{pk}}/E_{\mathrm{acc}}$ is not part of any objective function in the optimization problem [cf.~Eq.~(\ref{eq:OptimizationGoal})], these two geometric parameters can be omitted. A natural way to do this is to set $B = A$ and $b = a$, i.e., to consider circles instead of ellipses in the geometric parameterization of the cavity cross section [cf.~Fig.~\ref{fig:SingleCell}].
	
	\section{Constrained multi-objective optimization problem (CMOOP)}\label{sec:cmoop}
	Based on the information from Figs.~\ref{fig:wide-p2-main} and \ref{fig:narrow-p2-main},
	\[\frac{\partial f}{\partial R_{eq}},  \frac{\partial f}{\partial A},\text{ and } \frac{\partial f_{\text{TM}_{110}}}{\partial R_{eq}}\]
	need to be taken into account. Similarly, from the information shown in section~\ref{sec:frequency-fixing} (Figs.~\ref{fig:fix-freq-wide-p2-main} and \ref{fig:fix-freq-narrow-p2-main}),
	\[\frac{\partial f_{\text{TE}_{111}}}{\partial L}, \frac{\partial f_{\text{TE}_{111}}}{\partial R_i}, \frac{\partial f_{\text{TM}_{110}}}{\partial A},\text{ and } \frac{\partial f_{\text{TM}_{110}}}{\partial R_i} \]
	need to be considered as well. However, due to the influence of $A$ on $R_{eq}$ (Figs.~\ref{fig:fix-freq-wide-p2-main} and \ref{fig:fix-freq-narrow-p2-main}) and their geometric connection (Fig.~\ref{fig:SingleCell}), in order to decrease the computation cost the local sensitivities with respect to $A$ are omitted.
	
	Therefore, the constrained multi-objective optimization problem is the following [cf.~Eq.~(\ref{eq:OptimizationGoal})]

	\begin{equation}
		\begin{aligned}
		& \underset{R_i, L, A, a}{\text{min}}
		&& \Bigg(f-f_1,|f_1-f_2|,\frac{R}{Q}_{\perp1}+\frac{R}{Q}_{\perp 2}, \\
		& && -G\cdot \frac{R}{Q}, -R_i, \underbrace{\bigg|\frac{\partial f}{\partial R_{eq}}\bigg|}_{F_6}, \\
		& && \underbrace{\bigg|\frac{\partial f_{\text{TE}_{111}}}{\partial L}\bigg| + \bigg|\frac{\partial f_{\text{TE}_{111}}}{\partial R_i}\bigg|}_{F_7},\\
		& && \underbrace{\bigg|\frac{\partial f_{\text{TM}_{110}}}{\partial R_{eq}}\bigg| + \bigg|\frac{\partial f_{\text{TM}_{110}}}{\partial R_i}\bigg|\Bigg)}_{F_8}, \\
		& \text{subject to}
		&& f  = 400.79 \ \text{MHz}\ \text{ and } \ \alpha[^\circ] \geq 90. 
		\end{aligned}
		\label{eq:cmoop}
        \end{equation}

	It is implied that, for each point $\boldsymbol{I} = (R_i, L, A, a)$, the values of $B$ and $b$ will be set to $B=A$ and $b=a$, and that $f$ will be tuned to \unit[400.79]{MHz} using $R_{eq}$ as described in Section~\ref{sec:frequency-fixing}.
	
	\section{Constrained multi-objective optimization algorithm}\label{sec:cmooa}
	
	\subsection{Forward solver}\label{sec:forward-solver}
	In order to compute the values of the objective functions from Eq.~(\ref{eq:cmoop}) in a point $\boldsymbol{d} = (R_{eq}, R_i, L, A, B, a, b)$ [cf.~Eq.~(\ref{eq:d})] time-harmonic Maxwell's equations with perfectly electrically conducting (PEC) boundary conditions (BC) are solved in the evacuated axisymmetric RF cavity parameterized by $\boldsymbol{d}$. The mixed finite element method (FEM) leads to a generalized eigenvalue problem (GEVP) for each azimuthal mode number $m\in\mathbb{N}_0$~\cite{Arbenz2008,Chinellato2005}. 
	For monopole and dipole modes, $m = 0$ and $m=1$, respectively.
	If the cross section of the cavity is symmetric, as is the case for the single-cell elliptical cavity from Fig.~\ref{fig:SingleCell}, it is sufficient to solve time-harmonic Maxwell's equations for one half of it, once with PEC and once with perfectly magnetically conducting (PMC) BC on the cross-sectional symmetry plane (BC$_{\text{SP}}$).
	
	To compute the properties of the TM$_{010}$ mode, the smallest eigenpair of the GEVP corresponding to $m=0$ and PEC BC$_{\text{SP}}$ is found. This will be referred to as
	\[\text{MAXWELL}_{\boldsymbol{d}}\text{(}m\text{,BC$_{\text{SP}}$)} = \text{MAXWELL}_{\boldsymbol{d}}\text{(}0\text{,PEC)}.\]
	In order to compute the properties of the TM$_{110}$ and TE$_{111}$ mode, the smallest eigenpair of the GEVPs corresponding to $m=1$ and PEC and PMC BC$_{\text{SP}}$, respectively, is found. This will be referred to as
	\[\text{MAXWELL}_{\boldsymbol{d}}\text{(}1\text{,PEC) and MAXWELL}_{\boldsymbol{d}}\text{(}1\text{,PMC)}.\]
	
	\subsection{Optimization algorithm}\label{sec:optimization-algorithm}
	Since the minimizers of different objective functions are usually different points, the concept of dominance is used: a point $\boldsymbol{I}_1 = (R_{i,1}, L_1, A_1, a_1)$ dominates $\boldsymbol{I}_2$ if it is not worse in any of the objectives, and it is strictly better in at least one objective. A point is called Pareto optimal if it is not dominated by any other point. 
	
	Because of conflicting objectives, the ability of an evolutionary algorithm (EA) to escape local optima, its suitability for parallelization, as well as good results in the area of particle accelerator physics~\cite{Ineichen:13,PhysRevAccelBeams.22.054602,bazarov:05,hofler:13} a multi-objective EA~\cite{deb:09} is used to find an approximation of 
	the set of Pareto optimal points, even though other methods, such as particle swarm optimization~\cite{keeb:95,kara:05,hoss:09}, ant colony optimization~\cite{domc:96}, simulated annealing~\cite{kigv:83}, or artificial immune system~\cite{cati:02} exist.
	The basic steps of an EA are shown in Algorithm~\ref{alg:EA}. 
	
	In the case of the parameterization from Fig.~\ref{fig:SingleCell}, $N = 7$, the geometric parameters $d_1,\dots,d_7$ are [cf.~Eq.~(\ref{eq:d})] $R_{eq}$, $R_{i}$, $L$, $A$, $B$, $a$, and $b$, and $j = 1$, i.e., the frequency of the FM is tuned using $d_1 = R_{eq}$. A design point for the CMOOP in Eq.~(\ref{eq:cmoop}), also called an individual in the context of an EA, is \[\boldsymbol{I} = (d_2,d_3,d_4,d_6) = (R_{i}, L, A, B, a, b).\] The $M$ individuals ($M\in\mathbb{N}$) comprising the first generation are chosen randomly from the given intervals (Algorithm~\ref{alg:EA}:~line~\ref{alg:EA-initialize}). In this paper, in the case of the single-cell elliptical cavity, these intervals are given in Table~\ref{Tab:GeometryConstraints}. These $M$ individuals are then evaluated (\ref{alg:EA}:\ref{alg:EA-evaluate}), i.e., 
	the corresponding objective function values are computed, as shown in Algorithm~\ref{alg:evaluate}, in the following way.
	
	First, an $R_{eq}[\text{mm}] \in [325,375]$ [cf.~Table~\ref{Tab:GeometryConstraints}] is found such that the frequency of the FM of the corresponding cavity is $f(\boldsymbol{d}) = \unit[400.79]{MHz}$ (\ref{alg:evaluate}:\ref{alg:fix-frequency}). This is done using the zero-finding method TOMS~748~\cite{Alefeld1995}. Each evaluation of $f$ requires finding the smallest eigenpair of the GEVP (which will be referred to as `solving' the GEVP) corresponding to $m=0$ and PEC BC$_{\text{SP}}$.
	In case the cavity found this way is re-entrant, the second constraint is violated, so this individual is declared invalid and discarded from the population (\ref{alg:evaluate}:\ref{alg:discard}). Otherwise, the rest of the objective function values are computed (\ref{alg:evaluate}:\ref{alg:compute-objs}--\ref{alg:return-objs}). 
	
	In order to compute the properties of the TM$_{010}$ and TE$_{111}$, two GEVPs need to be solved (\ref{alg:evaluate}:\ref{alg:compute-objs}). In order to numerically compute the local sensitivity (i.e., the partial derivative) $\partial f_k/\partial d_l$ using the forward difference method, $f_k$ needs to be evaluated in the point $\boldsymbol{d}_{h,l} = (d_1,\dots,d_{l}+h,\dots,d_N)$, i.e., another GEVP needs to be solved for the appropriate azimuthal mode number $m_{f_k}$ and BC on the cross-sectional symmetry plane BC$_{\text{SP},f_k}$.
	
        \begin{algorithm}[H]
		\caption{Evolutionary algorithm\\Geometric parameters $d_1,\dots,d_N$, $f$ tuned using $d_j$, an individual is \[\boldsymbol{I} = (d_{i_1},\dots,d_{i_q}),\ \{i_1,\dots,i_q\}\subset \{1,\dots,N\}\setminus\{j\}.\]}\label{alg:EA}
		\begin{algorithmic}[1]
			\State random population of individuals $\boldsymbol{I}_1,\dots, \boldsymbol{I}_M$ \label{alg:EA-initialize}
			\State EVALUATE($\boldsymbol{I}_i, j$), $\forall i \in\{1, \dots, M\}$ \label{alg:EA-evaluate}
			\For {a predetermined number of generations} \label{alg:EA-cycle}
			\For {pairs of individuals $\boldsymbol{I}_i$, $\boldsymbol{I}_{i+1}$} \label{alg:EA-crossover1}
			\State crossover($\boldsymbol{I}_i$, $\boldsymbol{I}_{i+1}$), mutate($\boldsymbol{I}_i$), mutate($\boldsymbol{I}_{i+1}$) \label{alg:EA-crossover2}
			\EndFor
			\State for each new individual $\boldsymbol{I}_{new}$, EVALUATE($\boldsymbol{I}_{new}, j$) \label{alg:EA-evaluate-new}
			\State choose $M$ fittest individuals for the next generation \label{alg:EA-selector2}
			\EndFor
		\end{algorithmic}
	\end{algorithm}
	
        \begin{algorithm}[H]
		\caption{EVALUATE($\boldsymbol{I},j$)\\
			\textbf{In:} $j\in\{1,\dots,N\},\ \boldsymbol{I} = (d_{i_1},\dots,d_{i_q}),$ where \[\{i_1,\dots,i_q\}\subset \{1,\dots,N\}\setminus\{j\}\]
			\textbf{Out:} $\boldsymbol{F}(\boldsymbol{d} = (d_1,\dots,d_N))$ s.t.~$f(\boldsymbol{d}) = \unit[400.79]{MHz}$}\label{alg:evaluate}  
		\begin{algorithmic}[1]
			\State use $$\text{MAXWELL}_{\boldsymbol{d}}\text{(}0\text{,PEC)}$$ to compute $f(\boldsymbol{d})$ and TOMS~748~\cite{Alefeld1995} algorithm to find $$d_j\in[d_{lower},d_{upper}]$$ s.t. $$f(\boldsymbol{d})\text{[MHz]} - 400.79 = 0$$\label{alg:fix-frequency}
			\State if $\alpha(\boldsymbol{d})[^\circ] < 90$,
			{\bf return}\label{alg:discard}
			\State compute the properties of the dipole modes using
			$$\text{MAXWELL}_{\boldsymbol{d}}\text{(}1\text{,PEC) and MAXWELL}_{\boldsymbol{d}}\text{(}1\text{,PMC)}$$\label{alg:compute-objs}
			\State for each partial derivative $\partial f_k/\partial d_l$, with the corresponding $m_{f_k}$ and BC$_{\text{SP},f_k}$, and $\boldsymbol{d}_{h,l} = (d_1,\dots,d_{l}+h,\dots,d_N),$ compute
			$$\text{MAXWELL}_{\boldsymbol{d}_{h,l}}\text{(}m_{f_k}\text{,BC}_{\text{SP},f_k}\text{)}$$
			in order to numerically compute $\partial f_k/\partial d_l$ using the forward difference method\label{alg:partial-derivatives}
			\State {\bf return} $\boldsymbol{F}(\boldsymbol{d}) = (F_1(\boldsymbol{d}),\dots,F_n(\boldsymbol{d}))$\label{alg:return-objs}
		\end{algorithmic}
	\end{algorithm}
	
        Once the first generation of the EA is evaluated, a predetermined number of cycles is performed, each resulting in a new generation (\ref{alg:EA}:\ref{alg:EA-cycle}--\ref{alg:EA-selector2}). In each cycle, new individuals are created using crossover and mutation operators (\ref{alg:EA}:\ref{alg:EA-crossover1}--\ref{alg:EA-crossover2}), and their objective function values are again computed (\ref{alg:EA}:\ref{alg:EA-evaluate-new}) as shown in Algorithm~\ref{alg:evaluate}. The new generation is then chosen to comprise approximately $M$ fittest individuals (\ref{alg:EA}:\ref{alg:EA-selector2}).
	
	\section{Results}\label{sec:results}
	\subsection{Implementation and timings}\label{sec:ImplementationAndTiming}
	The implementation of the optimization algorithm from the previous section is based on a combination of a massively parallel implementation of an EA~\cite{Ineichen:13,PhysRevAccelBeams.22.054602}, written in C++ and parallelized using MPI, with the axisymmetric Maxwell eigensolver~\cite{moop}. For a point $\boldsymbol{d}$, a mesh of half of the cross section of the corresponding cavity is created using the Gmsh~\cite{gmsh} C++ API, and the resulting GEVPs are solved using the symmetric Jacobi--Davidson algorithm~\cite{Geus-JDSYM}. 
	
	In (\ref{alg:evaluate}:\ref{alg:fix-frequency}) a cheap solve on a coarse mesh is performed first, in order to compute a good approximation for $f$, $f_{\text{coarse}}$, (3--4 significant digits). Whenever $f_{\text{coarse}}$ is further away from \unit[400.79]{MHz} than a given value $\varepsilon$ (in this paper, $\varepsilon = \unit[1]{MHz}$),
	the value $f_{\text{coarse}}$ is used in order to speed up the computation. 
	Once the zero-finding method gets closer to \unit[400.79]{MHz} than $\varepsilon$, a more expensive solve, on a much finer mesh, is performed in order to compute five significant digits of $f$. As a zero-finding method, the Boost\footnote{https://www.boost.org} C++ library implementation of the TOMS~748 algorithm is used. Additionally, the zero-finding method is stopped as soon as five significant digits of $f$[MHz] match 400.79. When the search interval $[325,375]$ is first subdivided into three similarly-sized parts, (usually) using coarse solves, this entire approach requires, on average, only 2.2 fine solves in (\ref{alg:evaluate}:\ref{alg:fix-frequency}).
	The coarse solves use a mesh with around 20'000 triangles, and the fine ones around 500'000 triangles. 
	On one core of the Intel Xeon Gold 6150 a coarse solve (creating a mesh, computing five smallest eigenpairs and the objective function values) takes around \unit[2]{s}. A fine solve takes around \unit[95]{s} (\unit[18]{s} for creating a mesh, \unit[74]{s} for computing the eigenpairs, and \unit[3]{s} for computing the objective function values).
	Additionally, in (\ref{alg:evaluate}:\ref{alg:compute-objs}) two GEVPs on the same mesh need to be solved, so the mesh can be reused.
	Similarly, for each partial derivative that needs to be computed in (\ref{alg:evaluate}:\ref{alg:partial-derivatives}), the solves are grouped in such a way as to avoid remeshing (eigenproblems corresponding to the same point, i.e., the same cavity shape, are solved consecutively).
	
	To give an impression of the computation work for the entire optimization, using 108 processes on Euler~\Romannum{4} (Euler cluster\footnote{https://scicomp.ethz.ch/wiki/Euler} of ETH~Zurich, three Intel Xeon Gold 6150 nodes, each with 18 cores @ \unit[2.7]{GHz} and cache size \unit[24.75]{MB}) it took almost \unit[15]{h}~\unit[19]{min} to compute 60 generations with $M = 100$ (around 30\% of the evaluated individuals were discarded from the optimization).

	\subsection{Optimization results}\label{sec:optimization-results}
	
        \begin{figure*}[!htb]
	\centering
	\includegraphics[width=\textwidth,clip,trim={3cm 1.7cm 3cm 1.7cm}]{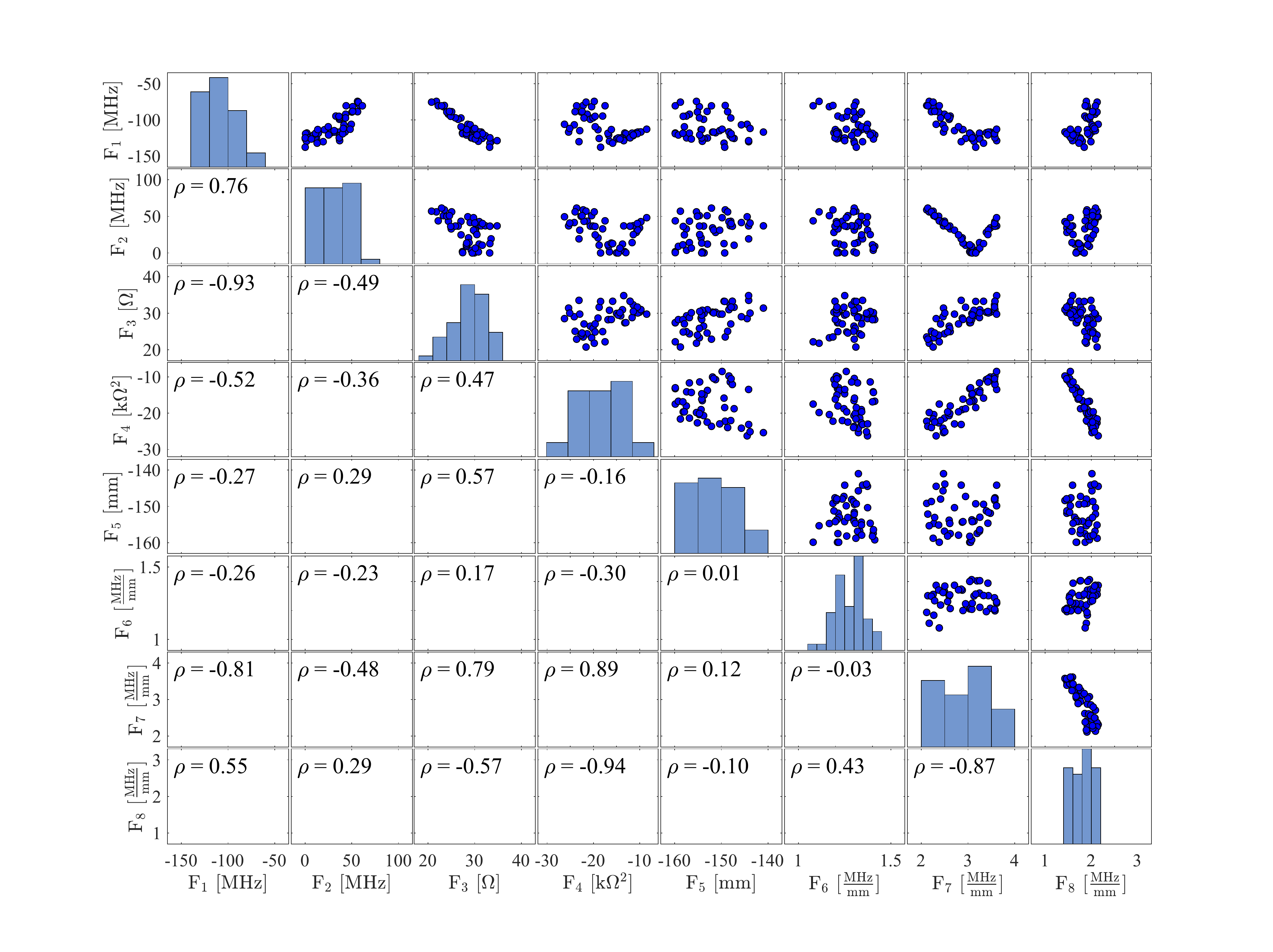}\hfill
	\caption{A scatter plot of the objective functions for the individuals in the last generation. Only individuals with $140 \leq R_i\leq 160$ and $200 \leq L \leq 400$ are shown in order to preserve the scale. The numbers below the diagonal are the correlation coefficients between pairs of objective functions.} 
	\label{fig:ObjfunPlot}
        \end{figure*}
	
The trade-off between the eight objective functions [cf.~Eq.~(\ref{eq:cmoop})] in the Pareto front approximation obtained in the 60-th generation of an optimization with $M = 100$ is illustrated in Fig.~\ref{fig:ObjfunPlot} using a scatter-plot matrix. The histogram at position $(i,i)$ (i.e., on the main diagonal) shows the distribution of the objective function $F_i$. The graph at position $(i,j)$ where $i < j$ (i.e., above the main diagonal) shows the values of $F_i$ ($y$ axis) and $F_j$ ($x$ axis) for the individuals in the last generation. The number in the $i$-th row and the $j$-th column where $i > j$ (i.e., below the main diagonal) is the correlation coefficient between $F_i$ and $F_j$.
$F_1$ and $F_2$ are positively correlated because $f_1$ (in most cases $f_1$ is $f_{\textrm{TE}_{111}}$) is more sensitive to the geometric changes than $f_2$, and the further it gets from $f$, the closer it gets to $f_2$. Improving $F_1$ and $F_2$ leads to a higher transverse impedance (larger $F_3$) and a more sensitive $f_{\textrm{TE}_{111}}$ (larger $F_7$). The V-shaped curve of $F_7$ vs $F_2$ is created because the rise in $F_2$  ($F_2 = |f_1-f_2|$) corresponds to cases where $f_{\textrm{TE}_{111}}$ is higher than $f_{\textrm{TM}_{110}}$. In order to simplify the damping of the dipole modes using coaxial HOM couplers, $F_2$ should be as close to zero as possible. A close-to-zero value of $F_2$ fixes $F_7$ to around \unit[3.1]{MHz/mm}. However, in choosing a point from the Pareto front approximation, a low importance can be assigned to $F_7$ because the transverse impedance of the ${\textrm{TE}_{111}}$ mode is much smaller than that of the ${\textrm{TM}_{110}}$ mode.  

All of the individuals, i.e., RF cavities, in the last generation are compared with four other storage ring single-cell cavities: CESR-B~\citep{CESRB}, HL-LHC~\cite{Roggen2015}, FCC$_{\text{HO18}}$~\cite{ZadehHOMSC18} and FCC$_{\text{IC18}}$~\cite{Marija18} in Table~\ref{Tab:nof-better}. Due to the conflicting nature of some of the objective functions, it is not feasible to surpass the other cavities in all objectives. However, all of these cavities are equally good in the Pareto sense.

	\begin{table}[h!]
		\centering
		\caption{The value in $i$-th row, $j$-th column is the number of distinct cavities that have at least $j$ objectives better than the cavity which corresponds to row $i$. The last (60-th) generation contains 95 distinct cavities.}
		\label{Tab:nof-better}
		\begin{tabular}{|c|c|c|c|c|c|c|c|c|}
			\hline 
			$j$ & 1 & 2 & 3 & 4 & 5 & 6 & 7 & 8 \\ \hline
			HL\hspace{1pt}-\hspace{1pt}LHC & 95 & 95 & 92 & 79 & 47 & 10 & 0 & 0 \\ \hline
			CESR\hspace{1pt}-\hspace{1pt}B & 95 & 95 & 92 & 68 & 40 & 9 &  0 & 0 \\ \hline
			FCC$_\text{HO18}$ & 95 & 94 & 78 & 51 & 19 & 1 & 0 & 0 \\ \hline
			FCC$_{\text{IC18}}$ & 95 & 95 & 93 & 86 & 35 & 5 & 0 & 0 \\ \hline
		\end{tabular}
	\end{table}

	\renewcommand{\arraystretch}{1.5}
	\begin{table*}[]
		\centering
		\caption{A comparison of a cavity on the Pareto front approximation, denoted FCC$_{\text{PR19}}$, with some storage ring single-cell cavities. The dimensions of CESR-B and HL-LHC cavities are scaled and tuned such that $f=\unit[400.79]{MHz}$. The cavities FCC$_{\text{HO18}}$ and FCC$_{\text{IC18}}$ are two other cavities, optimized with respect to a different set of objective functions, proposed for FCC by Zadeh and Kranj\v{c}evi\'c, respectively. The red and green color indicates that the highlighted cell has, respectively, a better and worse value than that of FCC$_{\text{PR19}}$.}
		
		\label{Tab:Cavity_Comparison}
		\begin{tabular}{|C{2.8cm}|C{1.5cm}|C{1.5cm}|C{1.5cm}||C{1.5cm}|C{1.5cm}|}
			\hline
			Variable                                                       &   FCC$_{\text{PR19}}$      & CESR-B        & HL-LHC & FCC$_{\text{HO18}}$ &FCC$_{\text{IC18}}$  \\ \hline
			$R_i$~[mm]                                                     &  153.704    &    150.0 & 150.0        & 156.0      &   141.614      \\ 
			$L$~[mm]                                                       &  274.199    &   300.0   &  280.0       & 240.0     &    292.54    \\ 
			$A$~[mm]                                                       &   53.582    &   103.750   &   104.0      &  70.0     &  103.54       \\ 
			$B$~[mm]                                                       &   53.582    &   103.750  &   104.0      &  70.0    &  127.521       \\ 
			$a$~[mm]                                                       &   36.6831   &   25.0   &   25.0      &   25.0    &   41.921       \\ 
			$b$~[mm]                                                       &   36.6831   &  25.0     &    25.0     &   25.0    &  45.812      \\ 
			$R_{eq}$~[mm]                                                  &  363.346    &  341.856  &  338.512       & 350.574    &    339.166        \\ \hline\hline
			QoI                                                            &   FCC$_{\text{PR19}}$       &   CESR-B      & HL-LHC & FCC$_{\text{HO18}}$ & FCC$_{\text{IC18}}$\\ \hline
			$f_{\text{TE}_{111}}$~[MHz]                                                    &  526.80     &    513.20  &  523.53 &  529.61 & 547.82   \\ 
			$f_{\text{TM}_{110}}$~[MHz]                                                    &  526.94     &   542.65   &  543.36 &  528.76 &  548.22       \\ 
			$R/Q$ [$\Omega$]                                               &   78.2     &    89.5    &  90.6   &  79.0     &  94.9       \\ 
			$R/Q_{\perp,{\text{TE}_{111}}}$ [$\Omega$]                                     &    3.2      &      5.5   &   4.6      &   2.3    &  5.1      \\ 
			$R/Q_{\perp 2,{\text{TM}_{110}}}$ [$\Omega$]                                     &   26.8     &     24.1   &   26.7      &    27.8   &  31.2       \\ 
			$\alpha$~[$^\circ$]                                            &  109.2     &     104.9   &   99.0      &   102.8    &  91.7       \\ 
			$\frac{E_{\mathrm{pk}}}{E_{\mathrm{acc}}}$~[-]                                   &    1.8     &     2.0   &   2.0      &   1.9    &  1.9       \\ 
			$\frac{B_{\mathrm{pk}}}{E_{\mathrm{acc}}}$~$\big[\frac{\unit{mT}}{\unit{MV/m}}\big]$     &    4.7     &     4.2   &    4.0        &   4.1    &   4.2      \\ \hline\hline
			Objective                                                      &   FCC$_{\text{PR19}}$      &  CESR-B      & HL-LHC & FCC$_{\text{HO18}}$ & FCC$_{\text{IC18}}$ \\ \hline
			$F_1$~[MHz]                                                     & -126.01     & \cellcolor{mygreen!15!white}{-112.42}    &  \cellcolor{mygreen!15!white}{-122.74}   & \cellcolor{red!10!white}{-127.97}      & \cellcolor{red!10!white}{-147.03}        \\ 
			$F_2$~[MHz]                                                     &    0.15     & \cellcolor{mygreen!15!white}{29.44}    &  \cellcolor{mygreen!15!white}{19.83}   & \cellcolor{mygreen!15!white}{0.85}   &     \cellcolor{mygreen!15!white}{0.40}       \\ 
			$F_3$~[$\Omega$]                                                &   30.0      & \cellcolor{red!10!white}{29.6}       &  \cellcolor{mygreen!15!white}{31.2}      & \cellcolor{mygreen!15!white}{30.1}      & \cellcolor{mygreen!15!white}{36.3}         \\ 
			$F_4$~[k$\Omega^2$]                                            &   -15.0     &  \cellcolor{red!10!white}{-21.8}       &   \cellcolor{red!10!white}{-21.3}     &  \cellcolor{red!10!white}{-15.5}    &   \cellcolor{red!10!white}{-21.3}      \\ 
			$F_5$~[mm]                                                     &  -153.704    &  \cellcolor{mygreen!15!white}{-150.0} & \cellcolor{mygreen!15!white}{-150.0} & \cellcolor{red!10!white}{-156.0} & \cellcolor{mygreen!15!white}{-141.614}      \\ 
			$F_6$~[$\frac{\unit{MHz}}{\unit{mm}}\big]$                          &    1.21    &\cellcolor{mygreen!15!white}{1.32}       &  \cellcolor{mygreen!15!white}{1.36}     & \cellcolor{mygreen!15!white}{1.31}     &  \cellcolor{mygreen!15!white}{1.39}       \\
			$F_7$~[$\frac{\unit{MHz}}{\unit{mm}}\big]$                          &    3.08    & \cellcolor{red!10!white}{2.71}       &  \cellcolor{red!10!white}{2.90}     & \cellcolor{mygreen!15!white}{3.16}     &   \cellcolor{mygreen!15!white}{3.20}     \\ 
			$F_8$~$\big[\frac{\unit{MHz}}{\unit{mm}}\big]$                          &    1.68    & \cellcolor{mygreen!15!white}{1.98}      &  \cellcolor{mygreen!15!white}{1.99}     & \cellcolor{mygreen!15!white}{1.79}     &   \cellcolor{mygreen!15!white}{1.92}     \\ \hline 
		\end{tabular}
	\end{table*}
	\renewcommand{\arraystretch}{1.25}

\begin{figure}[h!]
		\begin{tikzpicture}
		\node[anchor=south west,inner sep=0] (image) at (0,0){\includegraphics[clip,trim={0cm 8.5cm 0.5cm 7.8cm},width=1\columnwidth]{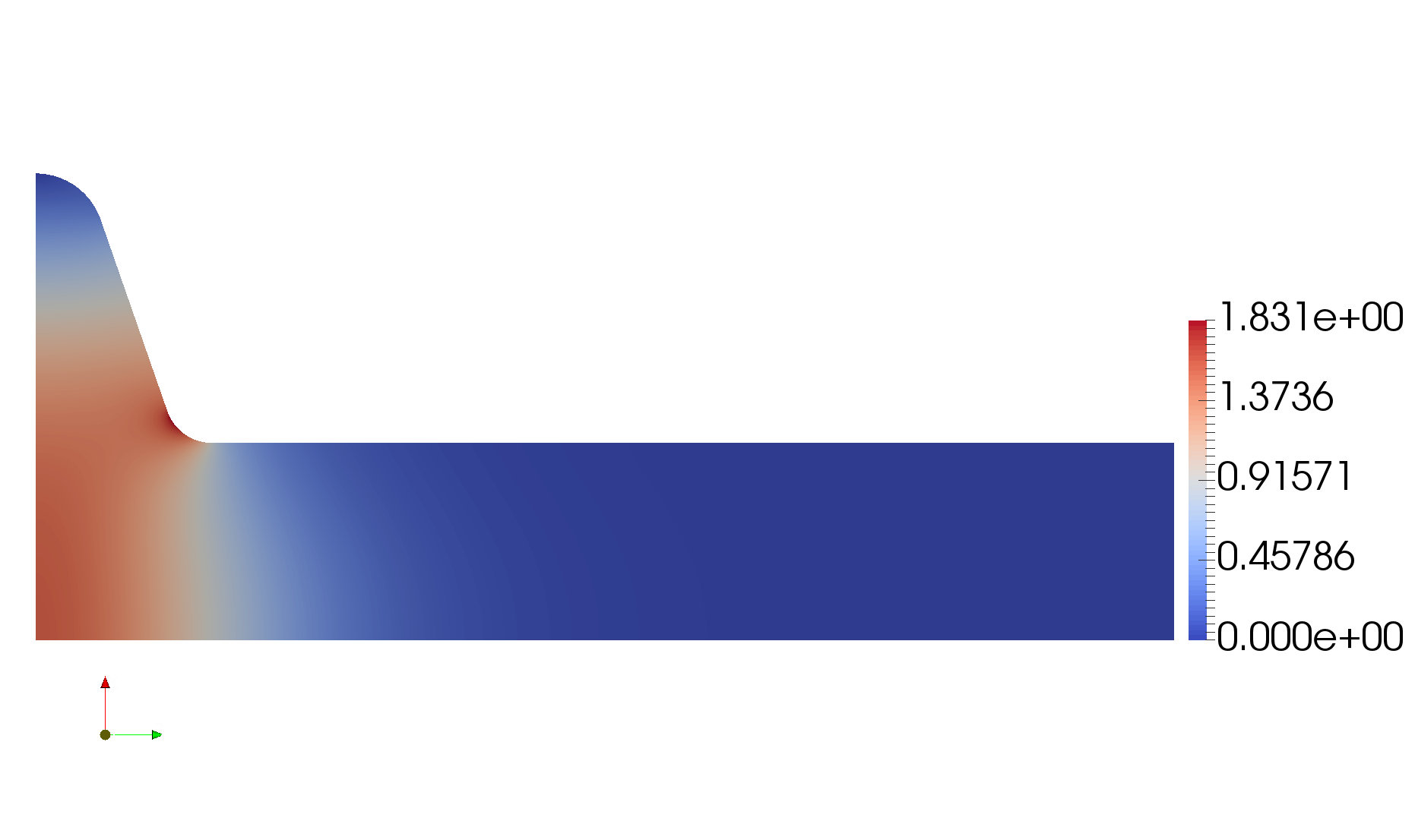}};
		\begin{scope}[x={(image.south east)},y={(image.north west)}]
		\node at (0.905,0.8) {$E/E_{\mathrm{acc}}$};
		\end{scope}
		\end{tikzpicture}
		\caption{The electric field of the FM in half of the chosen elliptical cavity, denoted FCC$_{\text{PR19}}$ in Table~\ref{Tab:Cavity_Comparison}.}
		\label{fig:elliptical-cavity}
	\end{figure}

The design variable and objective function values, as well as QoIs, for a chosen cavity from the last generation, called FCC$_{\text{PR19}}$, are given in Table~\ref{Tab:Cavity_Comparison}. The shape of FCC$_{\text{PR19}}$ is shown in Fig.~\ref{fig:elliptical-cavity}. The chosen cavity outperforms the other cavities in five or six objective functions. Since the design of FCC-ee-Z demands the cavities to be operated at low $E_{\mathrm{acc}}$~\cite[p.~120]{FCCCDR}, a small value of $G \cdot R/Q$ does not lead to a significantly large dynamical loss on the surface of cavity. Therefore, the value of $F_4$ is sacrificed in favor of other objective functions. The large iris radius of FCC$_{\text{PR19}}$ ensures that the dangerous higher order monopole modes are untrapped and propagate out of the beam pipes. 
Compared to other shown cavities, the frequency of the FM of FCC$_{\text{PR19}}$ is between 9\% and 15\% more robust against changes in $R_{eq}$. The local sensitivity of FCC$_{\text{PR19}}$ with respect to geometric parameters is given in Table~\ref{Tab:local-sensitivities}. These numbers confirm that the assumptions made in section~\ref{sec:cmoop} are valid for FCC$_{\text{PR19}}$.

        \begin{table}[t!]
		\centering
		\caption{Some of the local sensitivities for the RF cavity from Fig.~\ref{fig:elliptical-cavity}, denoted FCC$_{\text{PR19}}$ in Table~\ref{Tab:Cavity_Comparison}. The magnitude of other sensitivities is below \unit[0.7]{MHz/mm}.}
		\label{Tab:local-sensitivities}
		\begin{tabular}{|c|C{1.55cm}|C{1.3cm}|C{2.1cm}|}
			\hline
			Sensitivity &  $\partial f/\partial R_{eq}$ &  $\partial f/\partial A$ &  $\partial f_{\text{TE}_{111}}/\partial R_i$ \\ \hline
			Value~[\unit{MHz/mm}] & -1.21 & -1.15 & -2.62 \\ \hline
		\end{tabular}
		\begin{tabular}{|c|C{2.56cm}|C{2.56cm}|}
			\hline
			Sensitivity &  $\partial f_{\text{TM}_{110}}/\partial R_{eq}$ &  $\partial f_{\text{TM}_{110}}/\partial A$ \\ \hline
			Value~[\unit{MHz/mm}] & -1.34 & -1.00 \\ \hline
		\end{tabular}
	\end{table}
	
	\section{Generalization}\label{sec:generalization}
\begin{figure}[!tbh]
		\centering
		\includegraphics[width=\columnwidth,clip,trim={2cm 4cm 2cm 3.5cm}]{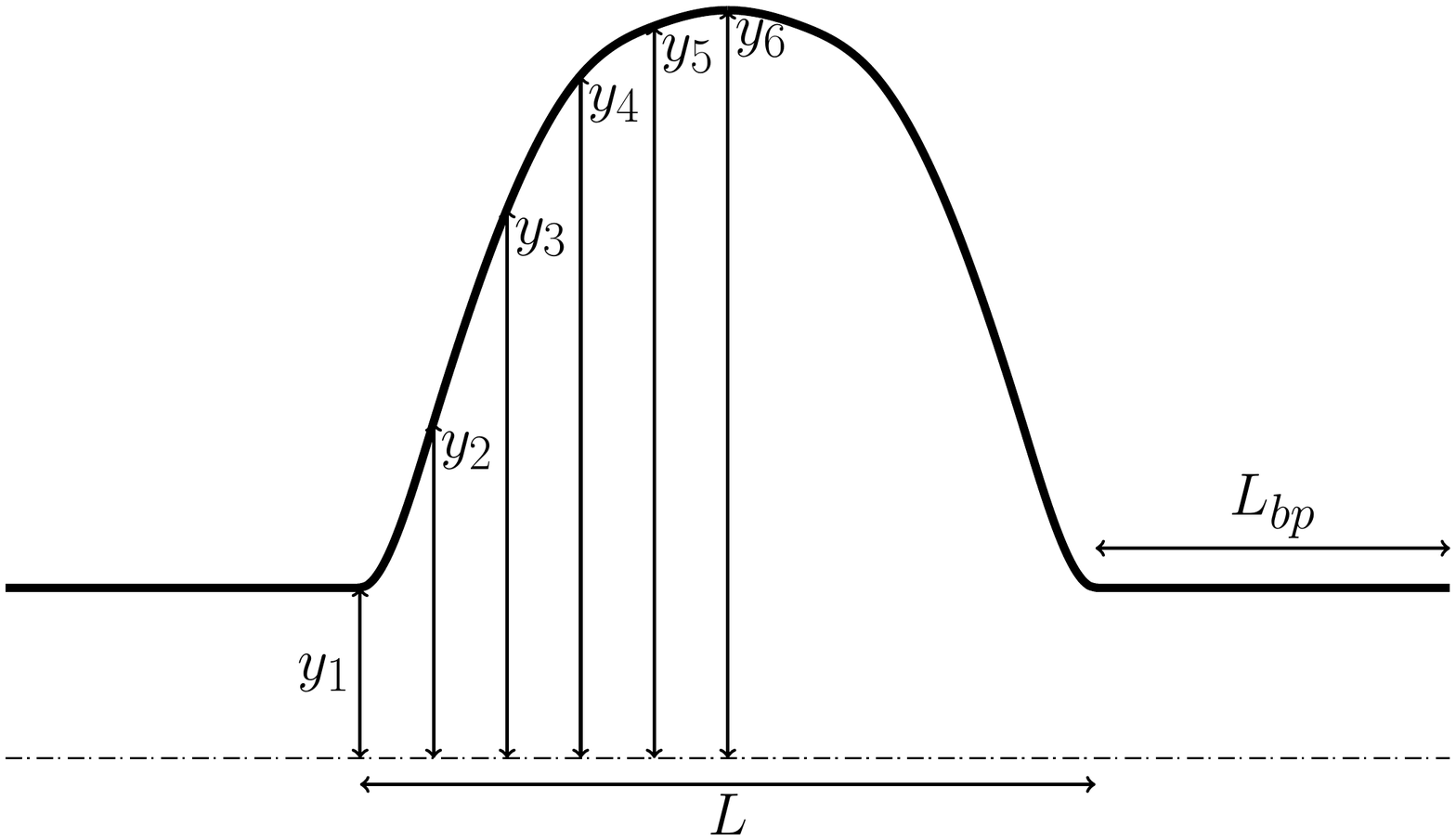}\hfill
		\caption{Parameterization of a single-cell spline cavity. The cross section is symmetric, and half of the curved part of the boundary is a complete cubic spline with $k$ equidistant knots and horizontal end slopes. 
		Preliminary computations showed interesting results for $k=6$.} 
		\label{fig:parameterization-spline}
	\end{figure}

	In this section the approach described in sections~\ref{sec:QoI}--\ref{sec:cmooa} will be generalized and applied to a single-cell cavity with a symmetric cross section and half of its boundary defined as a complete cubic spline with horizontal end slopes. Such a parameterization is shown in Fig.~\ref{fig:parameterization-spline} and referred to as `single-cell spline cavity'. 
	Related work includes the study of superconducting RF cavities whose boundary is a B\'{e}zier curve~\cite{Riemann:2012bha} or a non-uniform rational B-spline~\cite{georg2019uncertainty}.
        A summary of the approach described in sections~\ref{sec:QoI}--\ref{sec:cmooa} and its application to the single-cell spline cavity is given in Fig.~\ref{fig:generalization-flowchart}.
	
        \tikzstyle{method} = [rectangle, draw, fill=red!15!white, text width=9em,
	text centered, rounded corners, minimum height=2em, thick]
	\tikzstyle{conclusion} = [rectangle, draw, fill=blue!15!white, text width=15em,
	text centered, rounded corners, minimum height=2em, thick]                 
	\tikzstyle{l} = [draw, -latex', thick]
	
	\begin{figure}[h!]
		\begin{center}
                        \vspace*{25pt}\begin{tikzpicture}[scale=1]
			\node [conclusion] (parameterization) {Parameterization (Fig.~\ref{fig:parameterization-spline})};
			\node [method] (sensitivity) at ([shift={(0,-1.5)}] parameterization) {Sensitivity analysis (Fig.~\ref{fig:wide-p2-main-spline})};     
			\path [l] (parameterization)  -- (sensitivity);
			\node [conclusion] (f-params) at ([shift={(0,-3.25)}] parameterization) {Determine the most influential geometric parameter for $f$, $d_j$ ($d_j = y_4$)};  
			\path [l] (sensitivity)  -- (f-params) node[midway,right] {};
			\node [method] (sensitivity-freq-fix) at ([shift={(0,-5.2)}] parameterization) {Sensitivity analysis when $f$ is tuned using $d_j$ (Fig.~\ref{fig:fix-freq-wide-p2-main-spline})};
			\path [l] (f-params)  -- (sensitivity-freq-fix) node[midway,right] {};
			\node [conclusion] (all-params) at ([shift={(0,-7.35)}] parameterization) {Determine the influential design variables for the chosen QoIs ($f$ and $f_{\text{TM}_{110}}$) and define the CMOOP (Eq.~(\ref{eq:cmoopSplines}))};
			\path [l] (sensitivity-freq-fix)  -- (all-params) node[midway,right] {};
			\node [method] (ea-constraints) at ([shift={(0,-9.5)}] parameterization) {EA with constraint handling (Algorithms~\ref{alg:EA} and \ref{alg:evaluate})};
			\path [l] (all-params)  -- (ea-constraints);
			\node [conclusion] (pareto) at ([shift={(0,-11.3)}] parameterization) {Pareto front approximation and interesting individuals (Fig.~\ref{fig:spline-cavity})};
			\path [l] (ea-constraints)  -- (pareto);
			\path [l] (sensitivity) -| ([xshift=5mm] f-params.east) |- (all-params);
			\end{tikzpicture}
		\end{center}
		\caption{A summary of Sections~\ref{sec:QoI}--\ref{sec:cmooa}, applied to a different parameterization. The methods are red and the conclusions blue. The corresponding information for the single-cell spline cavity is referenced in parentheses.}
		\label{fig:generalization-flowchart}
	\end{figure}
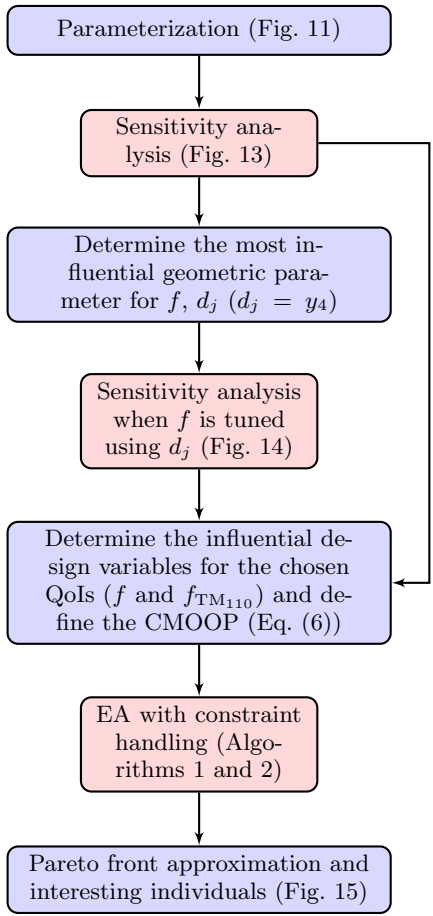
	
        \begin{table}[t!]
		\centering
		\caption{Intervals for the single-cell spline cavity.}
		\label{Tab:GeometryConstraintsSplines}
		\begin{tabular}{|l|c|c|c|c|c|c|c|}
			\hline
			Variable [mm] & $L$ & $y_1$ & $y_2$ & $y_3$ & $y_4$  & $y_5$ & $y_6$ \\ \hline
			Lower bound   & 240  & 145 & 150 & 220 & 270 & 320 & 340 \\ \hline
			Upper bound   & 380  & 160 & 230 & 280 & 330 & 345 & 360 \\ \hline
		\end{tabular}
	\end{table}
	
        \begin{figure}[h]
		\begin{center}
			\begin{minipage}[h!]{\columnwidth}
				\centering
				\includegraphics[width=\columnwidth,clip,trim={1.45cm 5.0cm 2.3cm 0cm}]{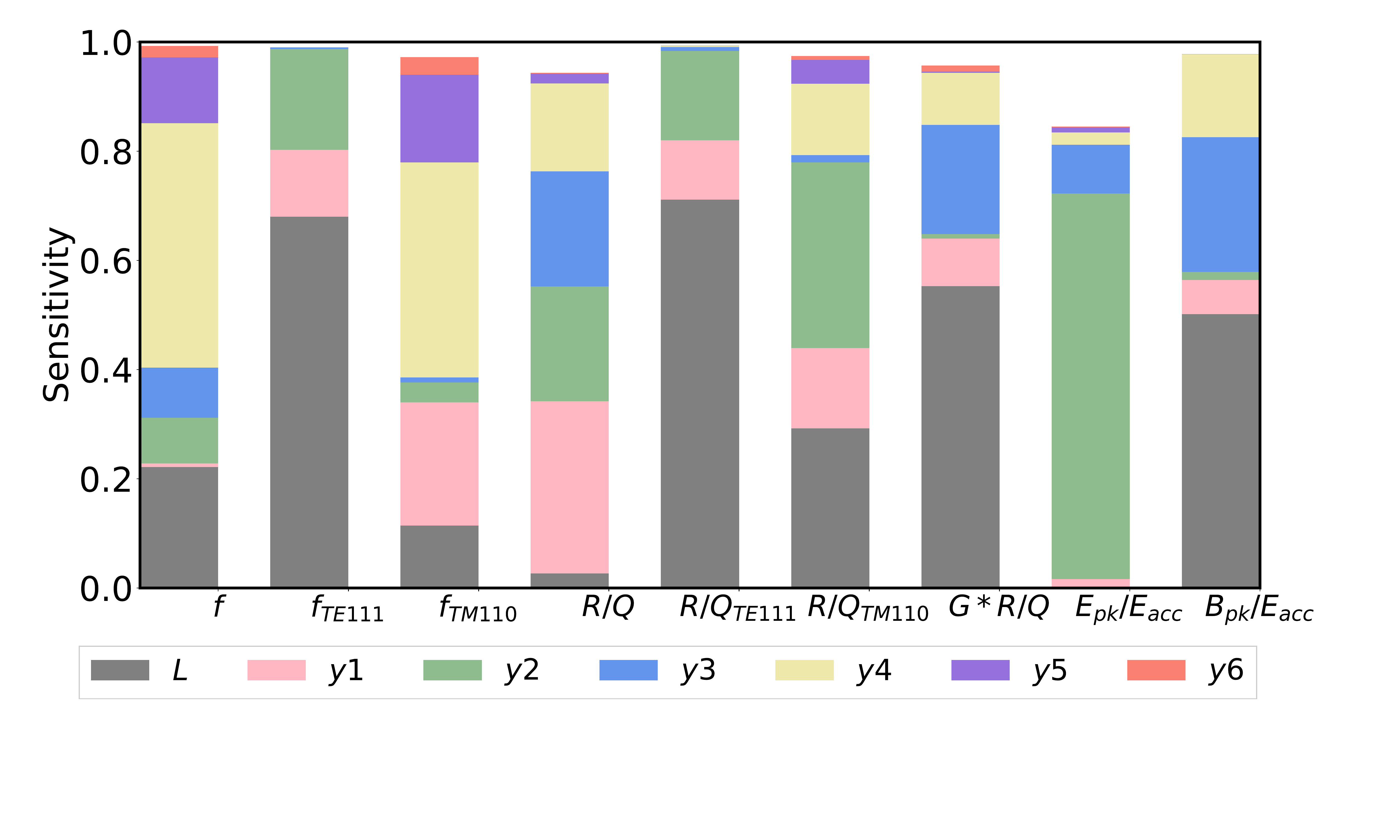}\hfill
				\caption{Main sensitivities. The intervals for the geometric parameters $L, y_1, \dots, y_6$ [cf.~Fig.~\ref{fig:parameterization-spline}] are given in Table~\ref{Tab:GeometryConstraintsSplines}. The polynomial degree is $p=2$, so the size of the sample is 216 [cf.~Table~\ref{Tab:nof-regression-pts}].} 
				\label{fig:wide-p2-main-spline}
			\end{minipage}
			
			\begin{minipage}[h!]{\columnwidth}
				\centering
				\includegraphics[width=\columnwidth,clip,trim={1.45cm 5.0cm 2.3cm 0cm}]{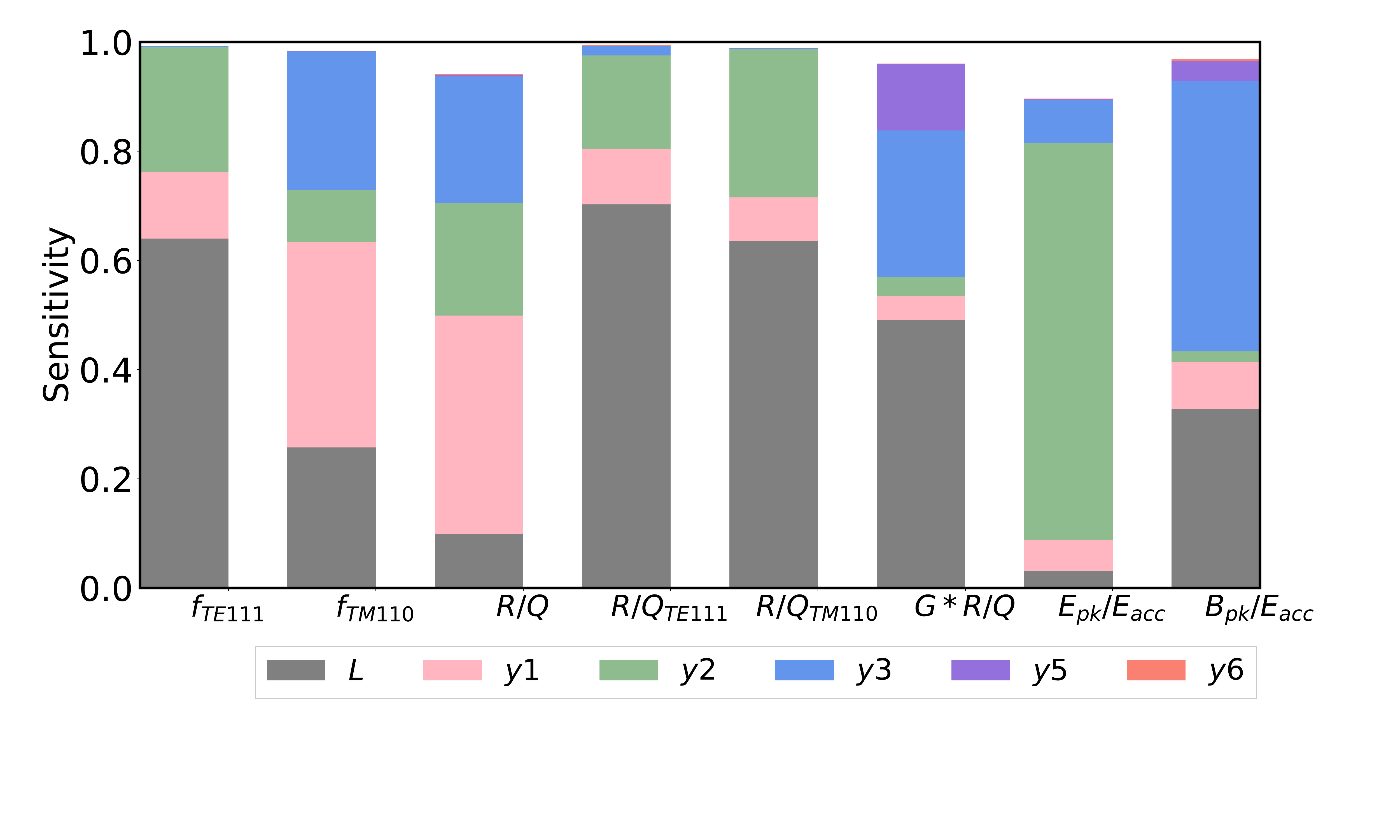}\hfill
				\caption{Main sensitivities. For a point $(L, y_1, y_2, y_3, y_5, y_6)$ [cf.~Fig.~\ref{fig:parameterization-spline}] inside the intervals from Table~\ref{Tab:GeometryConstraintsSplines}, $f$ is tuned to \unit[400.79]{MHz} using $y_{4}[\text{mm}]\in[270,330]$. The polynomial degree is $p=2$, so the size of the sample is 140 [cf.~Table~\ref{Tab:nof-regression-pts}].} 
				\label{fig:fix-freq-wide-p2-main-spline}
			\end{minipage}
		\end{center}
	\end{figure}
	
	Considering the intervals from Table~\ref{Tab:GeometryConstraintsSplines}, the main sensitivities are shown in Fig.~\ref{fig:wide-p2-main-spline}, where it can be seen that the geometric parameter with the greatest influence on the frequency of the FM is $y_4$. The main sensitivities in the case where, analogously to the use of $R_{eq}$ in section~\ref{sec:frequency-fixing}, $y_4$ is used to tune $f$ to \unit[400.79]{MHz} are shown in Fig.~\ref{fig:fix-freq-wide-p2-main-spline}. Note that the sensitivity of the spline cavity and the elliptical cavity cannot be compared as their shapes are parameterized differently. Based on the information from Fig.~\ref{fig:wide-p2-main-spline} and Fig.~\ref{fig:fix-freq-wide-p2-main-spline},
	\[\frac{\partial f}{\partial y_4}, \frac{\partial f_{\mathrm{TM}_{110}}}{\partial y_4} \text{ and } \frac{\partial f_{\mathrm{TE}_{111}}}{\partial L}, \frac{\partial f_{\mathrm{TM}_{110}}}{\partial y_1}\]
	need to be taken into account, respectively. Due to a rather low value of $R/Q_{\perp,{\text{TE}_{111}}}$ and the correlation observed between $F_2$ and $F_7$ in the previous section, we ignore the sensitivity of the TE$_{111}$ mode in the following optimization. Additionally, the influence of $y_6$ is small, so it can be omitted by, e.g., setting $y_6 = 350$ [cf.~Table~\ref{Tab:GeometryConstraintsSplines}]. 
	This leads to the constrained multi-objective optimization problem
	\begin{equation}
		\begin{aligned}
		& \underset{L, y_1, y_2, y_3, y_5}{\text{min}}
		&& \Bigg(f-f_1,|f_1-f_2|,\frac{R}{Q}_{\perp1}+\frac{R}{Q}_{\perp 2}, -G\cdot \frac{R}{Q}, \\
		& && \underbrace{-y_1}_{= F_5}, \underbrace{\bigg|\frac{\partial f}{\partial y_4}\bigg|}_{F_{G,6}}, \underbrace{\bigg|\frac{\partial f_{TM_{110}}}{\partial y_4}\bigg| + \bigg|\frac{\partial f_{TM_{110}}}{\partial y_1}\bigg|}_{F_{G,7}}\Bigg),\\
		& \text{subject to}
		&& f  = 400.79 \ \text{MHz.}
		\end{aligned}
		\label{eq:cmoopSplines}
        \end{equation}
	It is implied that, for a point $\boldsymbol{I} = (L, y_1, y_2, y_3, y_5)$, $y_6$ is $\unit[350]{mm}$ and $f$ will be tuned to \unit[400.79]{MHz} using $y_4$. 
	
\begin{figure}[h]
		\begin{tikzpicture}
		\node[anchor=south west,inner sep=0] (image) at (0,0){\includegraphics[clip,trim={0.5cm 7.6cm 0cm 3cm},width=1\columnwidth]{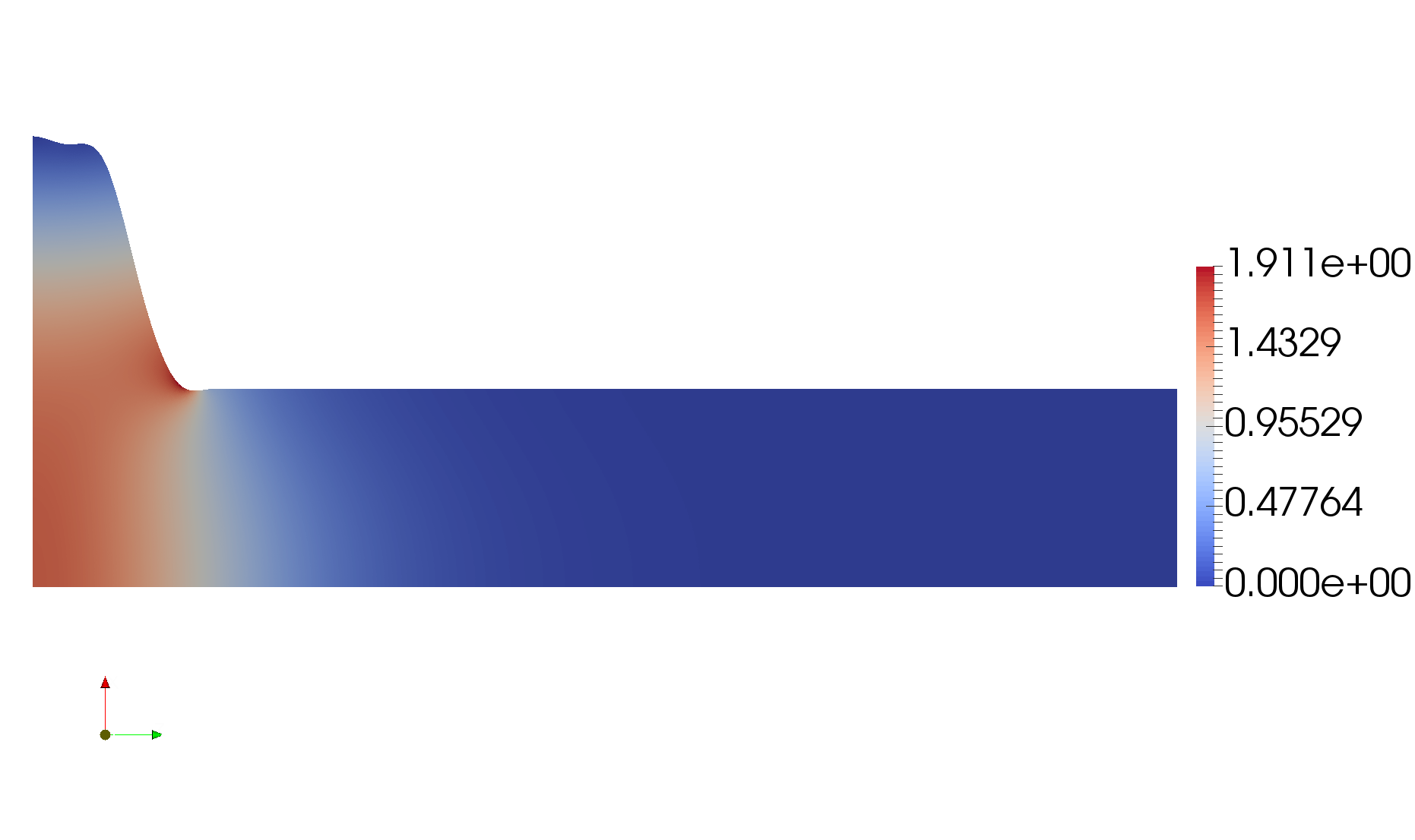}};
		\begin{scope}[x={(image.south east)},y={(image.north west)}]
		\node at (0.905,0.78) {$E/E_{\mathrm{acc}}$};
		\end{scope}
		\end{tikzpicture}
		\caption{The electric field of the FM in half of the chosen spline cavity.}
		\label{fig:spline-cavity}
	\end{figure}

	The optimization problem is solved as described in sections~\ref{sec:optimization-algorithm} and \ref{sec:ImplementationAndTiming}, and the shape of the chosen spline cavity shown in Fig.\ref{fig:spline-cavity}. The dip in the contour could be removed by, e.g., adding an additional constraint to the optimization problem. That is, however, not the goal of this section.
	
        The design variables, objective function values and QoIs for this cavity are given in Table~\ref{Tab:spline-additional-info}. The iris radius $y_1$, as well as the values of the objective functions and QoIs are very close to the corresponding values for the elliptical cavity found in section~\ref{sec:optimization-results} (FCC$_{\text{PR19}}$).
	
        \renewcommand{\arraystretch}{1.5}
        
        \begin{table}[b!]
	\centering
	\caption{Design variable and objective function values, as well as QoIs, for the spline cavity from Fig.~\ref{fig:spline-cavity}. The frequency of the FM is $f=\unit[400.79]{MHz}$, $F_5=-y_1$, $y_6 = \unit[350]{mm}$, and the smallest aperture width is denoted by $r_{min}$.}
	\label{Tab:spline-additional-info}
	\begin{tabular}{|C{2.8cm}|C{1.5cm}|}
		\hline
		 Variable  & Fig.~\ref{fig:spline-cavity} \\ \hline
		 $L$~[mm]    &  282.183  \\ 
		 $y_1$~[mm]  &  153.796  \\ 
		 $y_2$~[mm]  &  158.552  \\
		 $y_3$~[mm]  &  230.247  \\ 
		 $y_4$~[mm]  &  329.518  \\ 
		 $y_5$~[mm]  &  343.655  \\ 
		 $r_{min}$~[mm] & 152.311 \\ \hline\hline
		 QoI & Fig.~\ref{fig:spline-cavity}  \\ \hline
		 $f_{\text{TE}_{111}}$~[MHz]    &  530.13 \\ 
		 $f_{\text{TM}_{110}}$~[MHz]    &  530.21 \\ 
		 $R/Q$~[$\Omega$]     & 80.5  \\
		 $R/Q_{\perp \text{TE}_{111}}$~[$\Omega$] & 3.0  \\ 
		 $R/Q_{\perp \text{TM}_{110}}$~[$\Omega$]   & 27.8 \\ 
		 $\frac{E_{\mathrm{pk}}}{E_{\mathrm{acc}}}$~[-]   & 1.9  \\ 
		 $\frac{B_{\mathrm{pk}}}{E_{\mathrm{acc}}}$~$\big[\frac{\unit{mT}}{\unit{MV/m}}\big]$   & 4.8 \\ \hline\hline
                 Objective &  Fig.~\ref{fig:spline-cavity}  \\ \hline
		 $F_1$~[MHz] &  -129.34  \\ 
		 $F_2$~[MHz] &  0.08  \\ 
		 $F_3$~[$\Omega$] & 30.8 \\
		 $F_4$~[k$\Omega^2$] &   -15.8   \\ 
		 $F_5$~[$\frac{\unit{MHz}}{\unit{mm}}\big]$ &  -153.796  \\ 
		 $F_{G,6}$~[$\frac{\unit{MHz}}{\unit{mm}}\big]$ &  0.49 \\ 
		 $F_{G,7}$~$\big[\frac{\unit{MHz}}{\unit{mm}}\big]$ & 1.08 \\ \hline
	\end{tabular}
        \end{table}

	\renewcommand{\arraystretch}{1.25}
	
	\section{Conclusions}\label{sec:conclusions}
	In this paper an algorithm for solving constrained multi-objective RF cavity shape optimization problems was proposed and applied to the problem of optimizing the shape of the superconducting RF cavity for the FCC-ee-Z. The shape of the cavity was optimized with respect to both the properties of the fundamental mode and the first dipole band, focusing in particular on robustness against geometric perturbations. 
	
	In order to decrease the computation cost, the results of a global sensitivity analysis were used to reduce the search space and define the objective functions of interest. A good single-cell elliptical cavity was found, and the algorithm generalized and applied to a different type of cavity. The proposed algorithm and its implementation could also be applied to RF cavity shape optimization problems which take into account the properties of HOMs corresponding to arbitrary mode numbers.
	
	\begin{acknowledgments}
		This research was supported by the German Research Foundation (Deutsche Forschungsgemeinschaft, DFG) within the project RI 814/29-1.
		The computations were executed on the Euler compute cluster of ETH~Zurich
		at the expense of a Paul Scherrer Institut (PSI) grant.
		Matthias Frey helped with UQ and UQTk.
	\end{acknowledgments}

\end{document}